\documentclass[prd,aps,reprint,nofootinbib,superscriptaddress]{revtex4-2}

\usepackage{amsmath, amssymb, amsthm, mathtools, 
            esint, 
            hyperref, 
            mathrsfs, 
            graphicx} 

\usepackage{orcidlink} 

\newcommand{\Cornell}{\affiliation{Cornell Center for Astrophysics and Planetary Science, Cornell University, Ithaca, New York 14853, USA}}

\newcommand{\Caltech}{\affiliation{Theoretical Astrophysics 350-17, California Institute of Technology, Pasadena, CA 91125, USA}}

\newcommand{\AEI}{\affiliation{Max Planck Institute for Gravitational Physics (Albert Einstein Institute), Am M\"{u}hlenberg 1, Potsdam 14476, Germany}}

\newcommand{\CornellPhysics}{\affiliation{Department of Physics, Cornell University, Ithaca, NY, 14853, USA}}

\begin{document}

\title{Improved frequency spectra of gravitational waves with memory in a binary-black-hole simulation}

\author{Yitian Chen \orcidlink{0000-0002-8664-9702}} \email{yc2377@cornell.edu} \CornellPhysics \Cornell

\author{Michael Boyle \orcidlink{0000-0002-5075-5116}} \Cornell

\author{Nils Deppe \orcidlink{0000-0003-4557-4115}} \CornellPhysics \Cornell

\author{Lawrence E.~Kidder \orcidlink{0000-0001-5392-7342}} \Cornell

\author{Keefe Mitman \orcidlink{0000-0003-0276-3856}} \Caltech

\author{Jordan Moxon \orcidlink{0000-0001-9891-8677}} \Caltech

\author{Kyle C.~Nelli \orcidlink{0000-0003-2426-8768}} \Caltech

\author{Harald P.~Pfeiffer \orcidlink{0000-0001-9288-519X}} \AEI

\author{Mark A.~Scheel \orcidlink{0000-0001-6656-9134}} \Caltech

\author{William Throwe \orcidlink{0000-0001-5059-4378}} \Cornell

\author{Nils L.~Vu \orcidlink{0000-0002-5767-3949}} \Caltech

\author{Saul A.~Teukolsky \orcidlink{0000-0001-9765-4526}} \CornellPhysics \Cornell \Caltech

\begin{abstract}
	Numerical relativists can now produce gravitational waveforms with memory effects routinely and accurately. The gravitational-wave memory effect contains very low-frequency components, including a persistent offset. The presence of these components violates basic assumptions about time-shift behavior underpinning standard data-analysis techniques in gravitational-wave astronomy. This poses a challenge to the analysis of waveform spectra: How to preserve the low-frequency characteristics when transforming a time-domain waveform to the frequency domain.
	To tackle this challenge, we revisit the preprocessing procedures applied to the waveforms that contain memory effects. We find inconsistency between the zero-frequency limit of displacement memory and the low-frequency spectrum of the same memory preprocessed using the common scheme in literature. To resolve the inconsistency, we propose a new robust preprocessing scheme that produces the spectra of memory waveforms more faithfully. Using this new scheme, we inspect several characteristics of the spectrum of a memory waveform. In particular, we find a discernible beating pattern formed by the dominant oscillatory mode and the displacement memory. This pattern is absent in the spectrum of a waveform without memory. The difference between the memory and no-memory waveforms is too small to be observed by current-generation detectors in a single binary-black-hole event. Detecting the memory in a single event is likely to occur in the era of next-generation detectors.
\end{abstract}

\maketitle

\section{Introduction}

As gravitational-wave (GW) detectors become more sensitive, subtle physical effects in the signals become more detectable. Much of the GW literature gives the impression that GWs consist solely of oscillatory features. The typical pattern begins with a wave oscillating about zero in the inspiral phase; its frequency and amplitude gradually increase until peaking at the merger; and in the ringdown phase, the wave continues oscillating about zero while decaying away. These works include the first stable binary-black-hole (BBH) simulations \cite{Pretorius:2005gq, Baker:2005vv, Campanelli:2005dd}, the well-known first detection event \cite{LIGOScientific:2016aoc, LIGOScientific:2016vbw, LIGOScientific:2016emj, LIGOScientific:2016vlm} by the LIGO-Virgo-KAGRA Collaboration \cite{LIGOScientific:2014pky, VIRGO:2014yos, KAGRA:2020tym}, and various public waveform catalogs \cite{Mroue:2013xna, Jani:2016wkt, Healy:2017psd} as well as waveform models~\cite{Varma:2019csw, Pratten:2020ceb, Ossokine:2020kjp}. However, this picture is not complete for lack of one important physical effect: the gravitational-wave memory effect, or simply the \textit{memory}.

\subsection{Gravitational-wave memory}
\label{sec:intro_memory}

Memory in general refers to any permanent physical imprints to a spacetime left by the passage of a transient GW. Zel'dovich and Polnarev \cite{Zeldovich:1974gvh} first discovered this effect when investigating the gravitational scattering of two compact stars within the framework of linearized gravity. This effect is characterized by an enduring change in the GW strain between early and late times---in other words, a strain offset. This offset can cause a persistent relative displacement of two freely falling observers, so this particular memory effect is termed the \textit{displacement memory}. Being developed in linearized gravity, it is also called the \textit{linear memory}. Linear memory can be found in other gravitationally unbound systems, such as gamma-ray-burst (GRB) jets with matter emission \cite{Segalis:2001ns, Sago:2004pn} and supernovae with neutrino emission \cite{Epstein:1978dv, Turner:1978jj}. Modern nomenclature also refers to the linear memory as the \textit{ordinary memory} \cite{Bieri:2013ada}.

Displacement memory also exists in nonlinear general relativity, which was  discovered by several physicists independently \cite{Payne:1983rrr, Christodoulou:1991cr, Blanchet:1992br}. This nonlinear memory effect of a GW, also called the \textit{null memory} in modern literature \cite{Bieri:2013ada}, is sourced by the energy flux of the GW itself \cite{Wiseman:1991ss, Thorne:1992sdb}, so unlike linear memory, it can originate from gravitationally bound systems, or equivalently, from ``unbound'' systems where the emitted gravitons act as the ``unbound'' component \cite{Thorne:1992sdb}. Among these systems, only the most violent ones, e.g., compact object coalescences, can produce GWs in which nonlinear memory has a comparable strength to the usual oscillatory features \cite{Christodoulou:1991cr, Thorne:1992sdb}. Two additional nonlinear memory effects, \textit{spin memory} \cite{Pasterski:2015tva} and \textit{center-of-mass memory} \cite{Nichols:2018qac}, have been recognized lately. Unlike the displacement memory, these two new memories are more noticeable in the time integral of the GW strain and they are sourced by the angular momentum and center-of-mass flux in GWs.\footnote{There are memory effects in even higher order time integrals of the strain, identified in \cite{Grant:2021hga, Grant:2023jhd}.}.

In the past, connections have been made between the theories of memory and asymptotic symmetries.  The symmetry group for asymptotically flat spacetimes is the Bondi-van der Burg-Metzner-Sachs (BMS) group \cite{Bondi:1962px, Sachs:1962zza, Sachs:1962wk}, which extends the Poincar\'{e} group by enlarging the translation subgroup with an infinite-dimensional Abelian group of \textit{supertranslations} \cite{Newman:1966ub}. The displacement memory is closely related to the supertranslation and the associated charge, supermomentum~\cite{Bondi:1962px, Sachs:1962zza, Sachs:1962wk, Strominger:2014pwa, Flanagan:2015pxa}. More recently, the studies in \cite{Banks:2003vp, Barnich:2009se, Barnich:2010eb, Barnich:2010ojg, Barnich:2011mi, Campiglia:2014yka, Campiglia:2015yka} reexamined the symmetries for asymptotically flat spacetimes and proposed extensions to the BMS group. The extended BMS group contains super Lorentz transformations that generalize the regular Lorentz transformations \cite{Compere:2018ylh}. The magnetic- and electric-parity parts of the super Lorentz transformations are also called \textit{superrotations} and \textit{superboosts}. The spin and center-of-mass memory are respectively related to the superrotation (with superspin \cite{Flanagan:2015pxa} its conjugated charge) and superboost (with super center of mass \cite{Flanagan:2015pxa} the charge\footnote{A superspin and a super center of mass are collectively called a super angular momentum in \cite{Nichols:2018qac} or a superrotation charge in \cite{Hawking:2016sgy}.}). These theories, through a generalization of Noether’s theorem \cite{Wald:1999wa}, allow us to associate certain charges and fluxes as sources of the various memory effects at each instant of time, and therefore provide physicists the tools to extract time-dependent memory pieces from GWs~\cite{Strominger:2014pwa, Pasterski:2015tva, Flanagan:2015pxa, Mitman:2020pbt, Grant:2022bla}. These memory pieces contain nontrivial frequency contents that are potentially detectable and thus can act as a proxy for detection of memory.

There have been several recent studies on the detectability of the displacement memory. The fact that the displacement memory leads to a change in the proper separation between two freely falling objects may seem to offer a perfect detection scenario for a Michelson-interferometer-like GW detector. The current-generation detectors (including LIGO, Virgo, and KAGRA) are of this type, but they all have limited frequency bands that are designed for GW oscillations instead of memory-induced strain offsets, a relatively lower-frequency lower-amplitude feature. Thus, detecting the displacement memory is very unlikely in a single merger event for any current detectors. In fact, Refs.~\cite{Hubner:2019sly, Hubner:2021amk} found no evidence of the displacement memory in any individual GW event in the first two catalogs (GWTC-1 \cite{LIGOScientific:2018mvr} and GWTC-2 \cite{LIGOScientific:2020ibl}). There are two directions to improve the prospect of detecting the displacement memory. One is to rely on the next-generation detectors. The planned ground-based detectors Cosmic Explorer (CE) \cite{Reitze:2019iox} and Einstein Telescope (ET) \cite{Punturo:2010zz} will be able to observe the displacement memory arising from a single merger of two stellar-mass black holes \cite{Johnson:2018xly}. LISA \cite{LISA:2017pwj}, a future space-based detector, can also detect the displacement memory, but from supermassive BBHs \cite{Favata:2009ii, Islo:2019qht}. The other approach is to combine evidence from an ensemble of GW events \cite{Lasky:2016knh} in the current detectors, which is more promising in the near future. References~\cite{Hubner:2019sly, Boersma:2020gxx, Hubner:2021amk} all found that over a thousand BBH events are required to confirm the presence of the displacement memory. A comprehensive study using this approach that includes LIGO India \cite{Unnikrishnan:2013qwa} and CE can be found in \cite{Grant:2022bla}.

Regarding the spin memory, one physical effect is the differential time delay in two counter-orbiting light rays \cite{Pasterski:2015tva} after the passage of a GW, so the spin memory is potentially measurable in a Sagnac interferometer. Although the spin memory is characterized by an offset in the time integral of the GW strain, its time derivative (i.e., the rate of memory accumulation) exhibits a nontrivial signature in the strain. Thus, the spin memory may be detectable by CE and ET, using the technique of combining a population of BBH mergers \cite{Nichols:2017rqr, Grant:2022bla}.

Most research works on memory effects in the past two decades rely on some form of approximations of the memory. Without numerical relativity, traditional calculation of the memory can be done using post-Newtonian approximation (e.g., \cite{Favata:2008yd, Favata:2011qi, Blanchet:2013haa, Pasterski:2015tva, Nichols:2017rqr, Ebersold:2019kdc, Blanchet:2023sbv}). Nevertheless, both the displacement and spin memory grow most sharply near the merger, where the post-Newtonian theory breaks down. To obtain a realistic estimate of the memory through the merger phase, numerical relativity becomes necessary. Many studies calculated the effective memory from numerical waveforms that contain no memory \cite{Talbot:2018sgr, Mitman:2020bjf} or from models constructed based on such waveforms (including surrogate, numerical-relativity-calibrated effective-one-body, and phenomenological models; see \cite{Favata:2008ti, Favata:2009ii, Favata:2010zu, Talbot:2018sgr, Hubner:2019sly, Boersma:2020gxx, Grant:2022bla, Yoo:2023spi} for the memory calculation using these models). However, as the GW sources the memory, the memory also induces a portion of memory into itself. Computing memory from no-memory waveforms may fail to capture this ``memory of the memory''.\footnote{Reference~\cite{Talbot:2018sgr} investigated the importance of the memory-induced memory at different orders and found that the second-order memory is much smaller than the first-order memory by 3-4 orders of magnitudes, depending on the BBH configuration. Here, the first-order (second-order) memory is the memory calculated using the no-memory waveform (the no-memory waveform plus the first-order memory).}

It is not surprising that these studies did not use numerical waveforms with intrinsic memory, because such waveforms are largely unavailable. Indeed, most simulation codes extract asymptotic waveforms using extrapolation methods that fail to capture the memory. For example, Refs.~\cite{Mitman:2020pbt, Mitman:2020bjf} found that their waveforms, generated using the Spectral Einstein Code (\texttt{SpEC}) \cite{spec_skip} and extrapolated \cite{Iozzo:2020jcu} using the \texttt{scri} module \cite{scri_skip, Boyle:2013nka, Boyle:2014ioa, Boyle:2015nqa}, contain no displacement memory and roughly 50\% of the true spin memory. Note that almost the same extrapolation methods have been used to generate one of the largest public waveform catalogs, the Simulating eXtreme Spacetimes (SXS) catalog \cite{Mroue:2013xna, Boyle:2019kee}. Possible explanations for this imperfection of the extrapolation methods can be found in \cite{Favata:2010zu}.

Numerical waveform models with intrinsic memory do exist (e.g., \cite{Pollney:2010hs, Moxon:2021gbv}). They are quite rare, since they are generated using the methods of Cauchy-characteristic evolution (CCE) \cite{Bishop:1996gt, Winicour:2008vpn}, which are mathematically more complicated and thus harder to implement than the extrapolation schemes. Some examples of the CCE implementation can be found in \cite{Bishop:1997ik, Winicour:1999ba, Reisswig:2009us, Reisswig:2009rx, Handmer:2014qha, Barkett:2019uae}. More recently, an efficient CCE scheme \cite{Moxon:2021gbv} has been successfully implemented as a module of the \texttt{SpECTRE} code \cite{spectre_skip}. This scheme evolves waveform quantities including not only strain and news, but also all five complex Weyl components, which opens up the possibilities for fixing the BMS frame of a numerical waveform \cite{Mitman:2021xkq, Mitman:2022kwt}---i.e., the translation, supertranslation, rotation, and boost degrees of freedom. More relevant to our work, these accurate and robust CCE waveforms contain memory intrinsically, which enables extraction of precise displacement and spin memory \cite{Mitman:2020pbt}. 

\subsection{Signal processing}
\label{sec:intro_signal_processing}

Gravitational-wave data analysts rely primarily on optimal matched filtering to detect and characterize GW signals in noisy data streams~\cite{LIGOScientific:2019hgc, Allen:2005fk}. Time-domain signals are transformed to the frequency domain, where the stochastic noise background can be mitigated, and the matched filters can be efficiently applied for different times of arrival using an inverse transform. This efficient procedure relies on the assumption that the signal sufficiently settles down to the same value at both ends of the finite time domain, which is unfortunately violated by a raw waveform containing memory. By processing the signal appropriately, we can restore the validity of the procedure.

We begin by briefly reviewing matched filtering for a detector with noise characterized by the power spectral density (PSD) $S_n(f)$, and sensitivity limited to be between $f_{\min}$ and $f_{\max}$. Given two signals $s(t)$ and $h(t)$, we define their noise-weighted inner product as
\begin{equation}
	\label{eq:inner_product}
	\langle s | h\rangle = 4 \Re \int_{f_{\min}}^{f_{\max}} \frac{\tilde{s}(f)\, \tilde{h}^*(f)} {S_n(f)}\, df,
\end{equation}
where a tilde indicates a Fourier transform and the asterisk indicates the complex conjugate. In signal processing, Fourier transforms are discretized (discrete Fourier transforms, or DFTs), and in particular, they are generally implemented as \emph{fast} Fourier transforms (FFTs)~\cite{Cooley:1965zz}. In this equation, we think of $s$ as representing the detector data stream, and $h$ as a template waveform that may be present in the data.  We do not yet know the arrival time of the physical signal---if it is actually present---so we define the time-offset template waveform $h_{\delta t}(t) = h(t-\delta t)$, where $\delta t$ is to be determined. We then define the ``match'' between $s$ and $h$ as the maximum possible inner product for all values of $\delta t$:
\begin{equation}
	\label{eq:match}
	\mathrm{M}(s, h) = \max_{\delta t}\langle s | h_{\delta t}\rangle.
\end{equation}
Essentially, a large value of $\mathrm{M}(s, h)$ suggests the presence of a signal in the data stream $s$ that is well matched to the template $h$.

Crucially, we can convert the integral itself in Eq.~\eqref{eq:match} explicitly to an inverse Fourier transform. The ``shift theorem'' for Fourier transforms states that a time shift of a signal corresponds to a simple phase shift of its Fourier transform:
\begin{equation}
	\label{eq:shift_theorem}
	h(t) \to h(t-\delta t) \quad \Leftrightarrow \quad \tilde{h}(f) \to \tilde{h}(f)\, e^{-2\pi i f \delta t}.
\end{equation}
Thus, we may rewrite the match as
\begin{equation}
	\label{eq:matched_filter}
	\mathrm{M}(s, h) = \max_{\delta t} 4 \Re \int_{f_{\min}}^{f_{\max}} \frac{\tilde{s}(f)\, \tilde{h}^*(f)} {S_n(f)}\, e^{2\pi i f \delta t}\, df.
\end{equation}
We can then compute the integral as an FFT efficiently, making the maximization over $\delta t$ extremely fast.\footnote{Performing the integral as an FFT restricts the possible values $\delta t$ to a discrete set, which is not precisely what is called for when maximizing the inner product.  This is a form of ``scalloping loss'' or ``picket-fence effect''~\cite{Harris:1978yaq}.  If there are sufficiently many samples, the discretization may be fine enough for a given purpose, or the result may be used as an initial guess for more active optimization methods.}  This technique therefore forms the backbone of many GW data-analysis pipelines. For a finite-length time-domain signal, the ``time shift'' in the shift theorem is in fact ``circular'', i.e., the (infinite) periodic extension of the signal is shifted in time. However, this assumption is violated by a waveform containing memory, meaning that a naive application of Eq.~\eqref{eq:matched_filter} will rapidly lose sensitivity to the memory whenever $\delta t \neq 0$---that is, whenever the template does not happen to correctly guess the arrival time of the physical signal. In short, \emph{the presence of memory in a waveform invalidates the assumptions on which standard data-analysis techniques are based.} We will describe this problem in more detail in Sec.~\ref{sec:line_subtraction}. 

A DFT implicitly assumes that the finite-length signal is periodic, so the signal can be ``wrapped around'' and its two ends can be identified. Nevertheless, a BBH waveform is non-periodic, and more importantly, its two ends generally do not share the same value. This value mismatch, if left untouched, can lead to spurious contents in the frequency signal. Therefore, a numerical waveform must be \textit{preprocessed} in the time domain before being Fourier transformed. A common preprocessing scheme in the literature consists of tapering and padding. For a no-memory waveform, the inspiral oscillates around zero, and the ringdown signal settles down to zero (within numerical accuracy), so only the start of the inspiral signal needs to be tapered. In practice, the end of the ringdown signal is also tapered to remove any remaining numerical uncertainties. Padding is not strictly necessary in this case. One benefit of padding a no-memory waveform by zeros is to take advantage of the most efficient FFT algorithms. Now, suppose we have a memory waveform instead. Besides the inspiral part being tapered as usual, the ringdown part must be tapered as well due to the presence of the memory offset.\footnote{In this statement, we assume the displacement memory in the ringdown waveform does not settle down to zero.} Padding becomes essential if the ringdown signal is short. Indeed, a short ringdown signal requires a steep ringdown tapering that may adversely induce significant spectral leakages in the frequency spectrum. To avoid these artifacts, the end of the ringdown waveform should better be extended (i.e., padded by the end values) so that the ringdown tapering is sufficiently gradual.

There are many approaches to preprocessing that serve the purpose of reducing the effects of finite length and discrete sampling of a waveform signal, but a satisfactory scheme should produce a frequency spectrum as close as possible to that of a waveform that is arbitrarily long and arbitrarily finely sampled. In particular, the result should not depend sensitively on details of the preprocessing scheme. In this paper, we keep this goal in mind and revisit the preprocessing procedures that are applied to numerical waveforms with memory. We find that the low-frequency spectrum of displacement memory, preprocessed using the common scheme in literature, does not approach the zero-frequency limit (i.e., the behavior of the Fourier transform of a step function as $ f\to 0 $) \cite{Smarr:1977fy, Turner:1978jj, Thorne:1992sdb}, a feature of an infinitely long signal of the displacement memory.

To eliminate this inconsistency, we propose a new preprocessing scheme, which consists of padding, tapering, and an additional step we call \textit{line subtraction}.  The frequency spectrum of the displacement memory preprocessed using this new scheme correctly approaches the zero-frequency limit.  Moreover, the step of line subtraction also re-validates the application of the shift theorem in matched filtering.  We confirm the robustness of the new preprocessing scheme, by investigating how changing parameters in preprocessing can affect the resulting waveform spectrum.  As an application of this preprocessing scheme, we examine several characteristics of the frequency spectrum of a memory waveform. In particular, we find a clearly visible imprint of the displacement memory in the full waveform spectrum: a beating pattern formed by the displacement memory and the dominant oscillatory mode. Such an imprint is absent in a no-memory waveform spectrum. However, for a single general BBH merger event, the difference between the memory and no-memory waveform spectrum is too small to be observed by the current-generation GW detectors. Detecting memory in a single event is more likely to happen in the era of the next-generation detectors.

\subsection{Outline}
\label{sec:intro_outline}

The outline of this paper is as follows. In Sec.~\ref{sec:decomposition}, we briefly review a modern classification of memory effects in a gravitational waveform. This classification is based on different components of Einstein's equation in an asymptotically flat spacetime. In Sec.~\ref{sec:numerical_waveforms}, we describe the numerical extraction of the waveform studied throughout this paper, and the appearances of different memory pieces in this waveform. In Sec.~\ref{sec:preprocessing}, we introduce our new preprocessing procedure and compare it with other preprocessing procedures in literature. We also demonstrate the robustness of the new procedure by varying its parameters. In Sec.~\ref{sec:decomposition_spectrum}, we inspect the detailed structures of the frequency spectrum of the waveform that are preprocessed using the new scheme. In Sec.~\ref{sec:detecting_gw}, we briefly investigate the detection of GW signals in both the current- and next-generation detectors.  Finally, we summarize the results of this paper and consider future developments in Sec.~\ref{sec:conclusion}.

\section{Decomposition of strains} \label{sec:decomposition}

We can decompose the asymptotic waveform in two ways: the memory decomposition and the mode decomposition. The memory decomposition is based on asymptotic symmetries, while the mode decomposition is based on the angular dependence. In this section, we briefly review these two decompositions.

\subsection{Memory decomposition} \label{sec:memory_decomposition_theory}

Mitman \textit{et al.} \cite{Mitman:2020pbt} proposed a memory-based decomposition of a GW strain $h$. That is, using the notations in \cite{Mitman:2020bjf},
\begin{align}
	h = J_m + J_{\mathcal{E}} + J_{\widehat{N}} + J_{\mathcal{J}}. \label{eqn:memory_decomposition}
\end{align}
This relation holds at any time $t$\footnote{This is the retarded Bondi time, which is denoted as $u$ in most literature. We find using $t$ more natural in the discussion of preprocessing procedures, Fourier transforms and GW detection. Using $t$ instead of $u$ should cause no confusion, because there is only one time coordinate used in this paper.} along any angular direction $\theta^A=(\theta, \phi)$, and all quantities are assumed to be evaluated at future null infinity ($\mathscr{I}^+$). The four memory components $ J_\alpha $ ($ \alpha = m, \mathcal{E}, \widehat{N}, \mathcal{J} $) can be computed using only $h$ and the Weyl scalars (see \cite{Mitman:2020bjf} for the detailed expressions). 

This decomposition is deeply motivated by the physical representation of each component. The first term $J_m$ encodes the supermomentum charge and is closely related to the linear memory discovered by Zel'dovich and Polnarev \cite{Zeldovich:1974gvh}, which is also called the ordinary memory by Bieri and Garfinkle \cite{Bieri:2013ada}. The second term $J_{\mathcal{E}}$ corresponds to the GW energy flux and is associated with Christodoulou's nonlinear memory \cite{Christodoulou:1991cr}, also called the null memory by Bieri and Garfinkle \cite{Bieri:2013ada}. In other words, $J_{\mathcal{E}}$ represents the (nonlinear) displacement memory in the strain. The third term $J_{\widehat{N}}$ can be calculated from the superspin charge \cite{Flanagan:2015pxa}, while the fourth term $J_{\mathcal{J}}$ comes from the angular momentum flux of the GW. We also call the sum of the first two terms,
\begin{align}
	J^{(E)} = J_m + J_{\mathcal{E}},
\end{align}
the electric part of the strain, and the sum of the last two terms,
\begin{align}
	J^{(B)} = J_{\widehat{N}} + J_{\mathcal{J}},
\end{align}
the magnetic part of the strain. This is because $J^{(E)}$ and $J^{(B)}$ are derived from the electric- and magnetic-parity piece of the shear tensor used to define $h$ \cite{Flanagan:2015pxa, Mitman:2020pbt}. The time integral of $J^{(B)}$ is the spin memory. However, as we mainly deal with the strain itself instead of its time integral, we will also refer to $J^{(B)}$ as the ``spin memory''. Similar to $J_m$ and $J_{\mathcal{E}}$, the charge-associated $J_{\widehat{N}}$ and the flux-associated $J_{\mathcal{J}}$ can be regarded as the ordinary spin memory and the null spin memory. Thus, we will also call $J_m$, $J_{\mathcal{E}}$, $J_{\widehat{N}}$, and $J_{\mathcal{J}}$ the \textit{electric ordinary}, \textit{electric null}, \textit{magnetic ordinary}, and \textit{magnetic null} parts of the strain. Table~\ref{tbl:memory_components} summarizes the type and interpretation of these four memory components.

\begin{table}[t]
	\caption{Type and interpretation of the four memory components in the decomposition~\cite{Mitman:2020bjf} of the GW strain, Eq.~\eqref{eqn:memory_decomposition}.}
	\label{tbl:memory_components}
	\begin{ruledtabular}
		\begin{tabular}{ccc}
			Symbol & Type & Interpretation \\
			\hline 
			$J_m$ & Electric ordinary & Bondi mass aspect \\
			$J_{\mathcal{E}}$ & Electric null & Energy flux \\
			$J_{\widehat{N}}$ & Magnetic ordinary & Angular momentum aspect\\
			$J_{\mathcal{J}}$ & Magnetic null & Angular momentum flux
		\end{tabular}
	\end{ruledtabular}
\end{table}

\subsection{Mode decomposition}

The mode decomposition of the strain is
\begin{align}
	h(t, \theta, \phi) = \sum_{\ell = 2}^{L} \sum_{|m|\le \ell} h_{\ell m}(t) {}_{-2} Y_{\ell m}(\theta, \phi), \label{eqn:strain_modes}
\end{align}
where $ {}_{-2} Y_{\ell m}(\theta, \phi) $ are the spin-weighted spherical harmonics with the spin weight $s=-2$. In theory, the index $\ell$ increases without an upper bound, but in our numerical implementation, only those $\ell$ up to $L=8 $ are available. We introduce the notation $H_{\ell m}$ to represent the summands in Eq.~\eqref{eqn:strain_modes}. That is,
\begin{align}
	H_{\ell m}(t, \theta, \phi) = h_{\ell m}(t) {}_{-2} Y_{\ell m}(\theta, \phi).
\end{align}
To distinguish between $h_{\ell m}$ and $H_{\ell m}$ in the text, we refer to $h_{\ell m}$ as the $ (\ell, m) $ \textit{strain mode}, and $H_{\ell m}$ as the $ (\ell, m) $ \textit{strain component}. The concept of these $ (\ell, m) $ components turns out to be convenient when we later analyze the waveform fixed along a specific $ (\theta, \phi) $ direction, because these components simply add up to the overall strain in both the time domain (by definition) and frequency domain. 

The strain can also be expressed in terms of its two polarizations: $ h = h_+- ih_\times $. Here, $h_+$ and $h_\times$ represent the $+$ and $\times$ polarization components (which are both real) of the strain. Similarly, we can define the $+$ polarization of an $ (\ell, m) $ component as
\begin{align}
	H_{+,\ell m}(t, \theta, \phi) = \Re H_{\ell m}(t, \theta, \phi),
\end{align}
and the $\times$ polarization as
\begin{align}
	H_{\times,\ell m}(t, \theta, \phi) =-\Im H_{\ell m}(t, \theta, \phi).
\end{align}
They are related to $h_+$ and $h_\times$ by
\begin{align}
	h_+(t, \theta, \phi) &= \sum_{\ell = 2}^{L} \sum_{|m|\le \ell} H_{+,\ell m}(t, \theta, \phi), \label{eqn:mode_decomposition_plus} \\
	h_\times(t, \theta, \phi) &= \sum_{\ell = 2}^{L} \sum_{|m|\le \ell} H_{\times,\ell m}(t, \theta, \phi).
\end{align}

Similar to Eq.~\eqref{eqn:strain_modes}, each of the four components in the memory decomposition Eq.~\eqref{eqn:memory_decomposition} can be further decomposed into modes:
\begin{align}
	J_\alpha(t, \theta, \phi) = \sum_{\ell = 2}^{L} \sum_{|m|\le \ell} J_{\alpha, \ell m}(t) {}_{-2} Y_{\ell m}(\theta, \phi), \label{eqn:memory_modes}
\end{align}
where $\alpha$ can take the symbol $m$, $ \mathcal{E} $, $ \widehat{N} $, or $ \mathcal{J} $. For each mode, we have
\begin{align}
	h_{\ell m} = J_{m, \ell m} + J_{\mathcal{E}, \ell m} + J_{\widehat{N}, \ell m} + J_{\mathcal{J}, \ell m}. \label{eqn:memory_mode_decomposition}
\end{align}

\section{Numerical waveforms} \label{sec:numerical_waveforms}

We now introduce an example of a BBH simulation and illustrate the decomposition of its strain waveform as described in the previous sections. We will also use this waveform in the following sections to examine the effects of different options in preprocessing and to discuss the detectability of memory in GWs.

\subsection{Binary-black-hole simulation} \label{sec:bbh_simulation}

We numerically evolve an equal-mass, non-spinning, non-eccentric BBH system using the Spectral Einstein Code (\texttt{SpEC}) \cite{spec_skip} and list the simulation parameters in Table~\ref{tbl:bbh_param}. We simulate the BBH system at two different resolutions. The target truncation errors for the lower and higher resolutions are $ \sim 2\times 10^{-7} $ and $ \sim 5\times 10^{-8} $. Unless specified, the results in this paper are generated at the higher resolution.

\begin{table}[t]
	\caption{Parameters of the BBH simulation studied in this paper. The symbols $q$, $d_0$, $\omega_0$, $\dot{a}_0$, and $e$ stand for the mass ratio, initial coordinate separation, initial orbital frequency, initial rate of change of separation, and initial orbital eccentricity. The dimensionless spin vectors of the two BHs are denoted by $\vec{\chi}_{A,B}$. We choose the initial free data gauge to be the Gaussian superposition of two single BHs in their harmonic coordinates \cite{Cook:1997qc}. This gauge is called \textit{superposed harmonic} \cite{Varma:2018sqd} in \texttt{SpEC}.}
	\label{tbl:bbh_param}
	\begin{ruledtabular}
		\begin{tabular}{@{\hspace{3em}} cc @{\hspace{3em}}}
			Parameter & Value \\
			\hline 
			Initial free data & superposed harmonic \\
			$q$ & 1 \\
			$d_0$ & 20.7903$M$ \\
			$\omega_0$ & 0.009905$/M$ \\
			$\dot{a}_0$ & $-1.4\times10^{-5}$  \\
			$\vec{\chi}_{A,B}$ & $(0,0,0)$ \\
			$e$ & $ < 6\times10^{-6}$\\
		\end{tabular}
	\end{ruledtabular}
\end{table}

We compute the Weyl scalars ($ \Psi_{\text{0-4}} $) and the strain ($h$), all at $\mathscr{I}^+$, using the CCE module \cite{Moxon:2020gha, Moxon:2021gbv} in the \texttt{SpECTRE} code \cite{spectre_skip}. The CCE procedure requires specification of input data on two hypersurfaces. One is the inner boundary, and we choose it to be the world tube at the fixed coordinate radius $R_\text{world tube}=444M$,\footnote{In fact, we have tried four different world tubes. Their radii are equally spaced between $ R_{\min} = 101M $ and $ R_{\max} = 1130M $. Among these four world tubes, we pick the one with the second smallest radius, which renders the minimum violation of Bianchi identities. This practice is similar to those in \cite{Mitman:2020pbt, Mitman:2020bjf}.} where $M$ is the source-frame initial total Christodoulou mass of the BBH system. On this world tube, we specify quantities that are fully determined from the spacetime metric that is already provided in the \texttt{SpEC} BBH simulation. The other hypersurface is the initial null hypersurface, where we need to provide ``appropriate'' initial data that match the evolution history. We use the same practice as in \cite{Mitman:2020pbt, Mitman:2020bjf, Mitman:2021xkq, MaganaZertuche:2021syq, Mitman:2022kwt} and choose the following ansatz for the strain on the initial null hypersurface,
\begin{align}
	h(r, \theta, \phi) = \frac{A(\theta, \phi)}{r} + \frac{B(\theta, \phi)}{r^3},
\end{align}
The coefficients $A(\theta, \phi)$ and $B(\theta, \phi)$ are determined by matching the values of $h$ and $ \partial_r h $ to the world tube data. Here, $r$ is the radial coordinate in the Bondi-Sachs coordinates \cite{Bondi:1962px, Sachs:1962zza, Sachs:1962wk} and can be regarded as the distance from the BBH system \emph{in an asymptotically flat spacetime}. To account for cosmological redshift $z$, we can substitute $M$ and $r$ with the detection-frame initial total mass $M_{z} = M(1+z)$ and luminosity distance $d_{\mathrm{L}} = r(1+z)$.

We discard the first $ 1944M (=1500M + R_\text{world tube}) $ portion of the waveform, because it contains spurious GWs (often called ``junk'' radiation), which are present because the initial data fail to represent the true physical spacetime of a BBH evolution. The junk radiation shows up in both the Cauchy and characteristic systems. After we remove the junk stage, the remaining inspiral waveform has a length of $15373M$, which is about 38 inspiral orbits.  We shift the time axis of the waveform by choosing $t=0$ to be the moment when the $L^2$ norm $ \int |h|^2 \sin\theta d\theta d\phi $ reaches maximum. We regard $t=0$ as the merger time, and label the initial time $t_0$. We set the final time at $ t_f=200M $\footnote{The strain has indeed sufficiently settled down at $ t_f=200M $: The variation in the strain is at the level of $ 10^{-6} $ from $t=150M$ to $t=200M$.\label{fn:settle_last_50M}} and discard the portion of $ t>200M $ from the waveform. We fix the BMS gauge (including the center of mass) to the superrest frame of the remnant BH at $ t_f=200M $ using the BMS-charge-based method in \cite{Mitman:2022kwt}.\footnote{The LIGO-Virgo-KAGRA collaboration may expect to generate its waveform models in the post-Newtonian BMS frame, so it is perhaps slightly more natural to fix the simulated waveform in the post-Newtonian BMS frame~\cite{Mitman:2021xkq,Mitman:2022kwt}. However, because the algorithm presented in this work is independent of a BMS frame, we simply use the remnant superrest frame, to which it is easier to map~\cite{Mitman:2022kwt}.} 

We decompose the waveform obtained above into modes $h_{\ell m}$, and then decompose each mode into different memory contributions using Eq.~\eqref{eqn:memory_mode_decomposition}. Specifically, each waveform mode $h_{\ell m}$ is comprised of four contributions: the electric ordinary part $ J_{m, \ell m} $, the electric null part $ J_{\mathcal{E}, \ell m} $,\footnote{We fix the angular-dependent constant in the definition of $J_{\mathcal{E}}$, i.e., $\alpha$ in Eq.~(17b) of \cite{Mitman:2020bjf}, by simply equating both sides of Eq.~\eqref{eqn:memory_decomposition} (of this paper) at $t_f$. \label{fn:fix_J_E_const}} the magnetic ordinary part $ J_{\widehat{N}, \ell m} $, and the magnetic null part $ J_{\mathcal{J}, \ell m} $. We heavily rely on the packages \texttt{scri} \cite{scri_skip, Boyle:2013nka, Boyle:2014ioa, Boyle:2015nqa} and \texttt{sxs} \cite{sxs_skip} when dealing with the waveforms.

\subsection{Initial-offset removal} \label{sec:initial_offset_removal}

\begin{figure*}
	\centering	
	\begin{minipage}[b]{.483\linewidth}
		{\includegraphics[width=\linewidth]{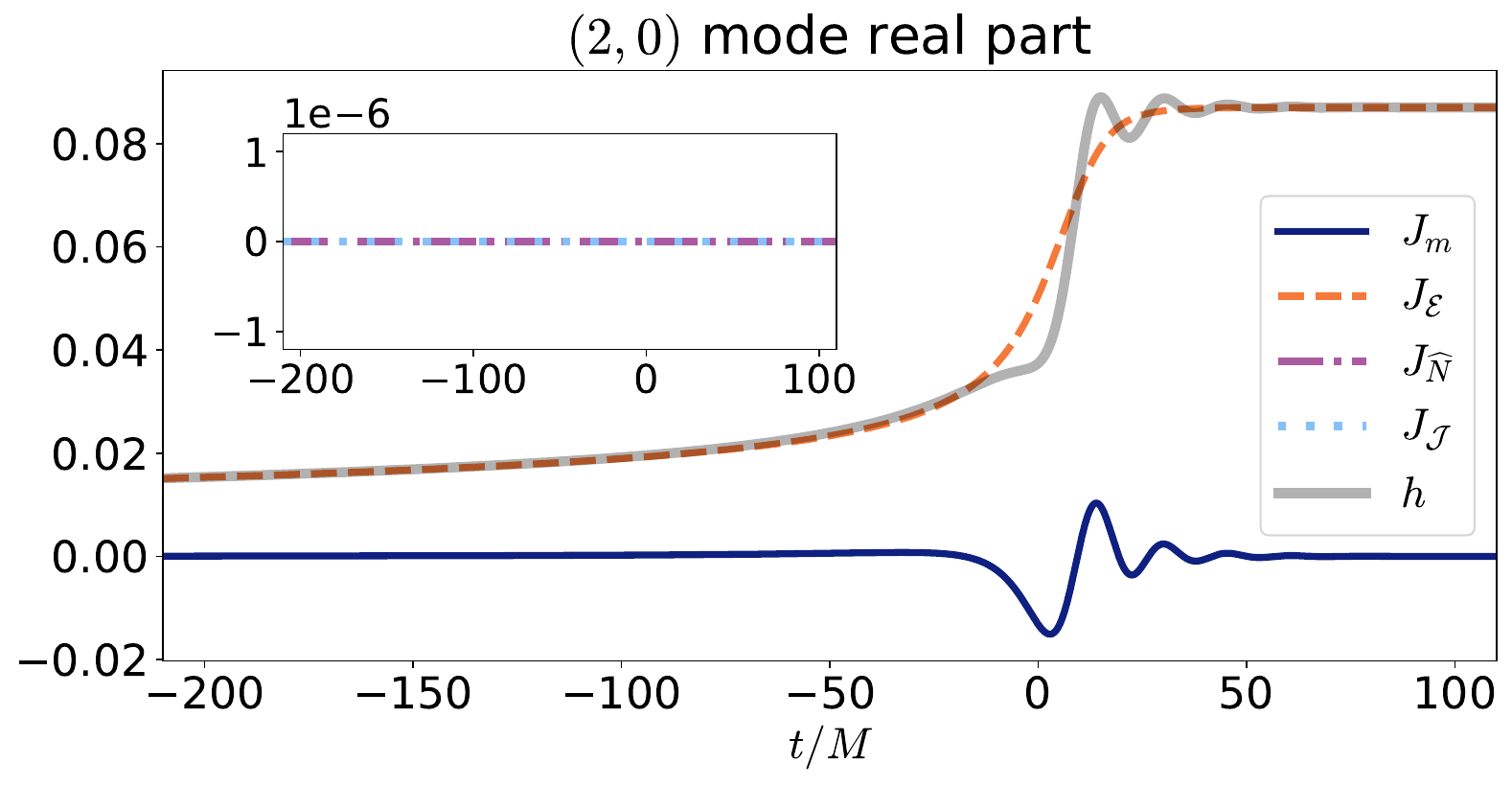}}
		{\includegraphics[width=\linewidth]{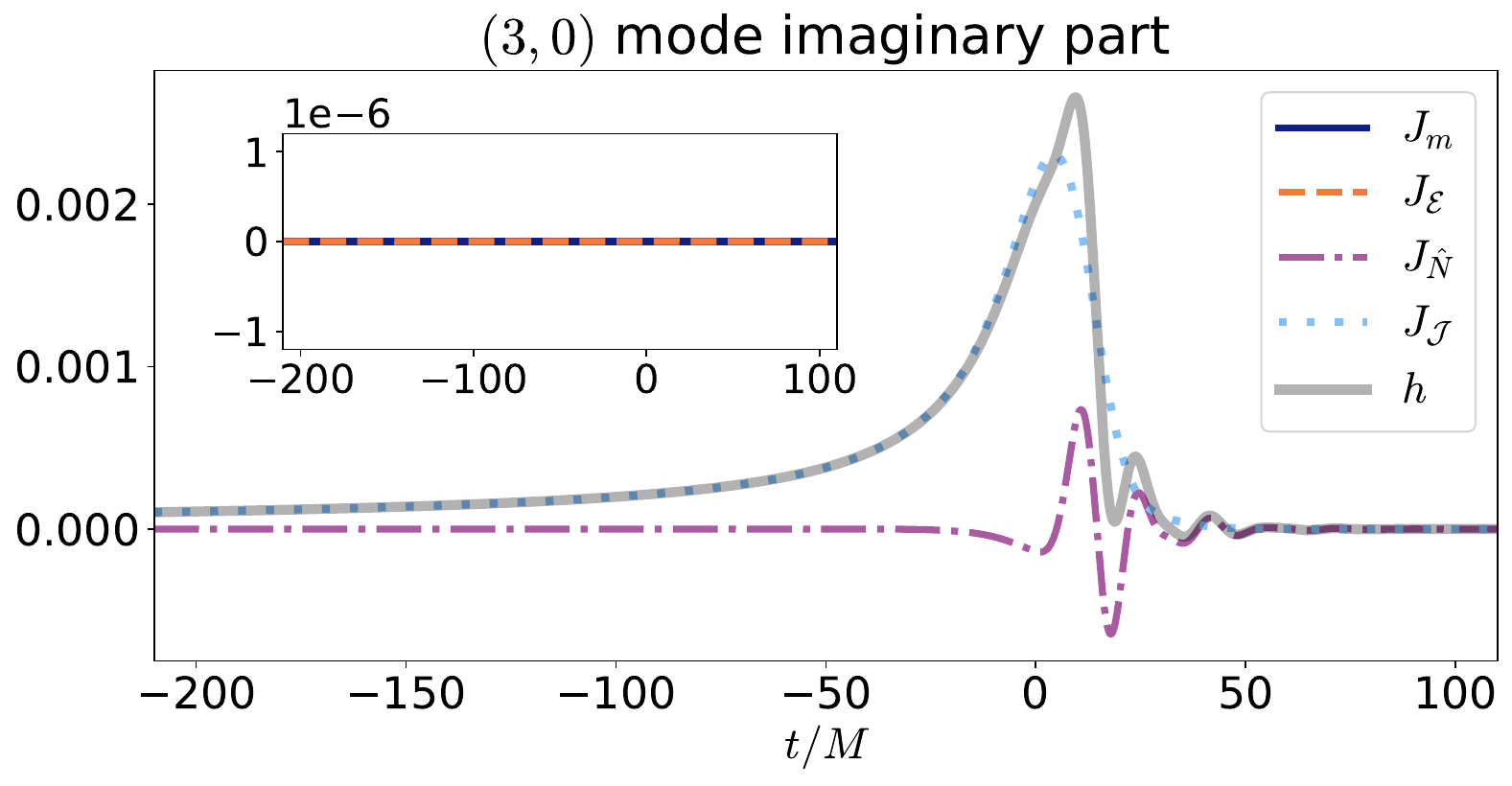}}%
	\end{minipage}
	\hfill
	\begin{minipage}[t]{.509\linewidth}
		{\includegraphics[width=\linewidth]{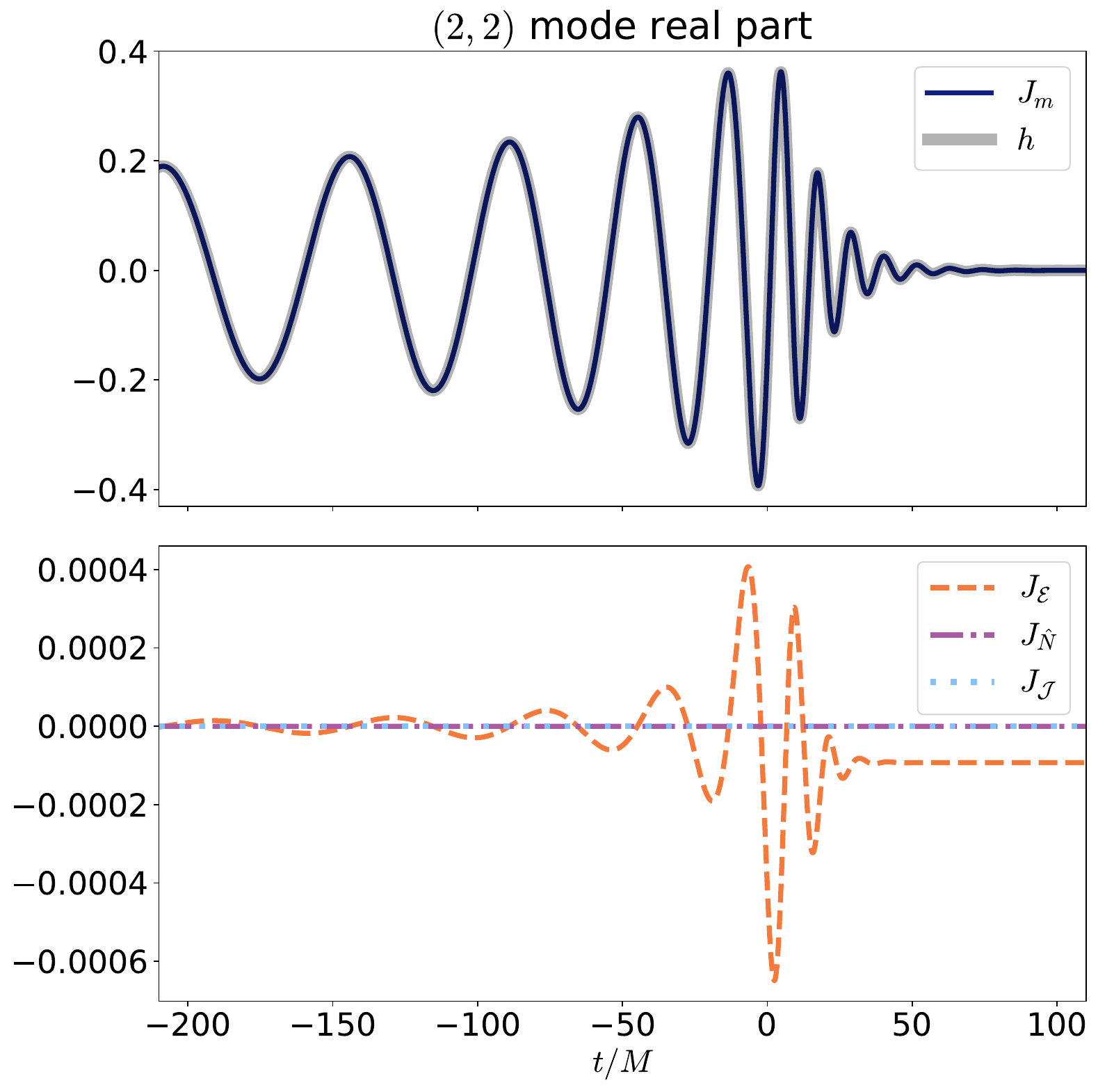}}%
	\end{minipage}
	\caption{Memory decomposition of the overall waveform for an equal-mass, non-spinning BBH. The strain $ h_{\ell m} $ (gray/thick), the electric ordinary part $ J_{m, \ell m} $ (blue/solid), the electric null part $ J_{\mathcal{E}, \ell m} $ (orange/dashed), the magnetic ordinary part $ J_{\widehat{N}, \ell m} $ (purple/dash-dotted), and the magnetic null part $ J_{\mathcal{J}, \ell m} $ (cyan/dotted) are functions of time. Top left: the real parts of these quantities for the $ (\ell=2, m=0) $ mode. Bottom left: the imaginary parts for $ (\ell=3, m=0) $. Right: the real parts for $ (\ell=2, m=2) $.}
	\label{fig:memory_modes}
\end{figure*}

The inspiral part of the waveform oscillates, but not necessarily about zero. Indeed, as the BMS gauge is now fixed to the superrest frame of the remnant BH (see the end of Sec.~\ref{sec:bbh_simulation}), all strain modes approach zeros at the end \cite{Mitman:2022kwt}. Due to the presence of the displacement memory, the early inspiral waveform should oscillate about a nonzero value. When we later taper the inspiral part of the waveform (see Sec.~\ref{sec:tapering}), not only the oscillations but also this nonzero offset will be tapered. Although adding or subtracting a constant from a time series does not affect the frequency spectrum at nonzero frequencies, tapering a constant function can induce spectral leakage (see Sec.~\ref{sec:no_removal_initial_offsets}). Thus, we would like to remove the initial offset from the waveform.

We mentioned in Sec.~\ref{sec:memory_decomposition_theory} that $ J_{\mathcal{E}} $, the electric null part, represents the displacement memory in the strain. We therefore choose the value of $ J_{\mathcal{E}, \ell m} $ at the initial time $t=t_0$ to be the initial offset of $ h_{\ell m} $, and subtract it from $ h_{\ell m} $. In other words, in the rest of this paper, we replace every waveform mode $ h_{\ell m}(t) $ (those generated in Sec.~\ref{sec:bbh_simulation}) by $ h_{\ell m}(t)- J_{\mathcal{E},\ell m}(t_0) $. The waveform quantities $ h_+ $, $h_\times$, $ H_{+,\ell m} $, and $H_{\times,\ell m} $ are also calculated based on these new $ h_{\ell m} $.

\subsection{Decompositions of strain} \label{sec:memory_decomposition_result}

Here, we investigate the following five waveform quantities: $h_{\ell m}$, $ J_{m, \ell m} $, $ J_{\mathcal{E}, \ell m} $, $ J_{\widehat{N}, \ell m} $, and $ J_{\mathcal{J}, \ell m} $. Note that the initial offsets of $h_{\ell m}$ and $ J_{\mathcal{E}, \ell m} $ have already been removed in Sec.~\ref{sec:initial_offset_removal}. Figure~\ref{fig:memory_modes} shows these quantities for three different modes.\footnote{We have checked that neither using a different world tube radius at the same resolution nor using the same world tube radius at the lower resolution would produce a visually different figure.} In all four panels, $ h_{\ell m} $, $ J_{m, \ell m} $, $ J_{\mathcal{E}, \ell m} $, $ J_{\widehat{N}, \ell m} $, and $ J_{\mathcal{J}, \ell m} $ are plotted in the following different styles: gray thick, blue solid, orange dashed, purple dash-dotted, and cyan dotted. The top left panel and the right two panels show the real parts of these quantities for the $ (\ell=2, m=0) $ and $ (\ell=2, m=2) $ modes. The bottom left panel shows the imaginary parts for the $ (\ell=3, m=0) $ mode. The memory decomposition of these modes is not new and has been studied in \cite{Mitman:2020pbt, Mitman:2020bjf}. These curves closely resemble those in the middle row of Fig.~2 in \cite{Mitman:2020pbt}.

In the top left panel of Fig.~\ref{fig:memory_modes}, we see that $J_{\mathcal{E}}$ is the dominant contributing factor to the $ (2,0) $ mode.\footnote{The $ (2,0) $ strain mode $ h_{2,0} $ is real, because $ h_{\ell,-m} = (-1)^\ell \bar{h}_{\ell, m} $ for a non-precessing BBH configuration \cite{Kidder:2007rt, Boyle:2014ioa}.\label{fn:real_h20}} Representing the (nonlinear) displacement memory, $J_{\mathcal{E}}$ indeed exhibits an offset between early and late times. In the bottom left panel, we see that the $ (3,0) $ mode\footnote{The $ (3,0) $ strain mode $h_{3,0}$ is imaginary. See Footnote~\ref{fn:real_h20}.} is mainly described by $J_{\mathcal{J}}$, (the time derivative of) the non-oscillatory spin memory. Though there is a small difference between the left- and right-end values of the $J_{\mathcal{J}}$ curve in this graph, we expect this difference to vanish if we could extend the curve to $ t =-\infty $.\footnote{At $ t =-\infty $ or $ t =\infty $, the spacetime is non-radiative. The news is zero, so by Eq.~(17d) of \cite{Mitman:2020bjf}, $J_{\mathcal{J}}$ vanishes.} In the top right panel, we observe that the $ (2,2) $ mode is faithfully represented by the electric ordinary part $ J_{m,22} $ (only the real parts are shown). We would also like to emphasize that the electric null part $ J_{\mathcal{E},22} $ makes up $\sim 0.1$\% of $h_{22}$ and that $ J_{\mathcal{E},22} $ not only contains a persistent offset but also oscillations. This is demonstrated in the bottom right panel of Fig.~\ref{fig:memory_modes}.

\section{Preprocessing} \label{sec:preprocessing}

The waveform $h$ and its various components ($H_{\ell m}$, $H_{+, \ell m}$, $h_+$, etc) are functions of $t$, $\theta$, and $\phi$. In this section, we consider the direction $ (\theta, \phi) $ fixed and use $a(t)$ to denote a general time-dependent waveform quantity. The numerical waveform $a(t)$ is a time series. Let $ \{ a_i | i = 0,...,n-1 \} $ be the time series representation of $a(t)$, and $ \{ t_i | i = 0,...,n-1 \} $ the underlying time array equally spaced by $ \Delta t $. We transform a numerical waveform from the time domain to the frequency domain using a discrete Fourier transform (DFT), more specifically, a fast Fourier transform (FFT) \cite{Cooley:1965zz}.

As discussed in the introduction, the two ends of a general numerical waveform do not match, and this mismatch, if not handled before applying the DFT, can lead to spectral leakage, which is a spurious effect that mixes spectral contents among frequency bins. To suppress this leakage, it is essential to preprocess the waveform before applying the DFT. In this section, we first discuss several tools for preprocessing: padding (Sec.~\ref{sec:pad}), tapering (Sec.~\ref{sec:tapering}), and line subtraction (also Sec.~\ref{sec:line_subtraction}). In Sec.~\ref{sec:standard_procedure}, we outline the standard preprocessing procedure used in this paper. In Sec.~\ref{sec:compare_other_procedures}, we illustrate this procedure's effectiveness compared to alternative preprocessing schemes with examples.

\subsection{Padding} \label{sec:pad}

Padding a discrete time series involves increasing the number of data points by simply prepending or appending data points before applying the DFT, and is a common technique in all domains of signal analysis, including within the GW community~\cite{Hinder:2017sxy, Varma:2018mmi, Boyle:2019kee, Boersma:2020gxx, Grant:2022bla, Gasparotto:2023fcg}. It serves two major purposes. One is to take advantage of the most efficient FFT algorithms, especially when the time series is padded to a length of a power of 2. We are not concerned about this purpose, because the FFT is never a bottleneck of the computational expense in our work. The other purpose is to narrow the frequency bin width and increase the \textit{display} resolution of the frequency spectrum. This does not mean the spectrum is better resolved \textit{physically}. Indeed, zero padding in the time domain is equivalent to a sinc-function-like interpolation in the frequency domain \cite{Smith1_skip, Engelberg1_skip}, so zero padding does not provide extra information. However, for a binary-merger signal, there is a better choice of padding than merely padding by zeros. Unlike a general time series, a GW signal from a general binary merger has a special feature: It finally settles down to a constant. Padding the right end of a sufficiently decayed waveform by that final (typically nonzero) value introduces additional physical information to the waveform. This padding is more favorable than zero padding, because it renders a \emph{more physically realistic} spectrum. Note that the resulting padded waveform is still far from complete, for lack of a full inspiral signal. 

If the ringdown waveform has not yet fully decayed at the end, the unsettled damped oscillations (which are quasinormal modes \cite{Teukolsky:1972my, Teukolsky:1973ha, Press:1973zz, Teukolsky:1974yv}) can contaminate the padding step. Directly applying a constant padding to the right end leads to a discontinuity in the first (or higher) order derivative and consequently an artificial $1/f^2$ (or faster) asymptotic high-frequency feature in the spectrum. One can remedy this by transitioning the last part of the waveform to a constant. The standard $ C^\infty $ transition function is
\begin{align}
	\mathcal{T}(t, t_1, t_2) = \begin{cases}
		0 & t \le t_1, \\
		\left[1+\exp \left(\frac{t_2-t_1}{t-t_1}+\frac{t_2-t_1}{t-t_2} \right)\right]^{-1} & t_1 < t < t_2, \\
		1 & t \ge t_2,
	\end{cases} \label{eqn:transition_function}
\end{align}
where $t_1$ and $t_2$ are two timesteps to be specified. A possible transitioning-to-a-constant procedure is to replace the final---say---$20M$ portion of the waveform, $a(t)$, by
\begin{align}
	a(t_f-20M) + \int_{t_f - 20M}^{t} \frac{da}{ds}(s) \mathcal{T}(s, t_f - 20M, t_f) ds, \label{eqn:transition_to_constant}
\end{align}
for $ t \in [t_f-20M, t_f] $.\footnote{This has been implemented as the \texttt{transition\_to\_constant} function previously in the \texttt{scri} package \cite{scri_skip, Boyle:2013nka, Boyle:2014ioa, Boyle:2015nqa}.} Although the new final value at $t=t_f$ depends on the behavior of $a(t)$, it must lie in between the minimum and maximum in the last $20M$ of $a(t)$. In this paper, we apply Eq.~\eqref{eqn:transition_to_constant} to our strain,\footnote{Our waveform ends at $t_f=200M$ and has sufficiently decayed (see Footnote~\ref{fn:settle_last_50M}). We checked that the difference between the spectra with and without the smooth transition Eq.~\eqref{eqn:transition_to_constant} is negligible. We still apply this transition for completeness and future reference.} and then pad it by the final value to an additional length of $20000M$ by default.

\subsection{Tapering}
\label{sec:tapering}

Having improved the physical realism of the signal by padding the end of the signal, we now look back to the beginning of the signal.  Recall that, as mentioned in the introduction, the finite DFT incorporates an implicit assumption that the underlying signal is periodic, with period equal to the length of the data.  This is certainly not true for a binary-merger signal, which will typically have been oscillating for a much longer time than the data includes; the data will turn on quite suddenly while the inspiral GW signal is fairly large.  The assumption of periodicity then implies that there is a discontinuity in the signal, which leads to a spectral interference error resembling the usual Gibbs phenomenon---oscillations in the spectrum's amplitude and phase~\cite{Harris:1978yaq}.

The standard method for dealing with this problem is to ``taper'' or ``window'' the signal, smoothly turning it on at the beginning and off at the end~\cite{Harris:1978yaq, Gabriele1_skip}.  The GW signal from a binary merger has a somewhat unusual structure: it chirps upward in frequency as the binary inspirals, and then transitions to exponential decay as the remnant rings down.  Ignoring the memory for a moment, this leads to fairly clean separation between frequency components at different times, with the frequency contents of the signal at early times turning up mostly at low frequencies.  This, in turn, leads to a natural choice of tapering functions that rise quickly from $0$ amplitude at the beginning, and then pass the rest of the signal through unchanged.

Most tapering functions used in signal processing do not have such a region where the signal passes through unchanged, but there are two common tapering functions that do.  One is the classic bump window (used in, e.g., \cite{McKechan:2010kp, LIGOScientific:2011hqo, Varma:2018mmi, Boyle:2019kee, Islam:2022laz, Grant:2022bla}), which is sometimes known as the Planck-taper window\footnote{``Bump functions'', such as this one, have long been used in, e.g., discussions of partitions of unity in differential geometry \cite{Gallier1_skip, Tu1_skip}.  It was referred to as the \textit{Planck-taper window} by \cite{McKechan:2010kp} because of a claimed similarity to Planck's law. However, we find the resemblance unconvincing and so the invocation of Planck to be a misattribution. Therefore, we refer to this window as a bump window.}  and can be expressed as the product of two transition functions of the form Eq.~\eqref{eqn:transition_function}:
\begin{align}
	\mathcal{W}(t, t_1, t_2, t_3, t_4) = \mathcal{T}(t, t_1, t_2) [1-\mathcal{T}(t, t_3, t_4)], \label{eqn:bump_window}
\end{align}
where $ t_1 < t_2 < t_3 < t_4 $.  This is precisely $0$ before $t_1$ and after $t_4$, precisely $1$ between $t_2$ and $t_3$, and transitions smoothly between those values.  The Tukey, or cosine-tapered, window is similarly shaped but only once differentiable.  Generally, all classic window functions~\cite{Harris:1978yaq} are symmetric, going to zero---or close to it---at both ends.

As mentioned in Sec.~\ref{sec:pad}, the final value of the signal may not be $0$, so tapering to zero at late times would not be helpful.  Instead, we use an asymmetric function, to taper only the \emph{beginning} of the signal.  For simplicity, we choose the transition function $\mathcal{T}$ of Eq.~\eqref{eqn:transition_function}.  The function should begin to rise from zero---at $t_1$ in Eq.~\eqref{eqn:transition_function}---as soon as the waveform has predominantly useful physical content. Because our waveform starting from $t=t_0$ has already excluded the ``junk radiation'' (see the end of Sec.~\ref{sec:bbh_simulation}), we simply choose $t_1=t_0$.  The value of $t_2$ representing the time at which the function reaches its full value of $1$ should be chosen based on other criteria.  We discuss these criteria for our example waveform in Sec.~\ref{sec:vary_taper_length}.  We choose $t_2 = t_{\text{cross-30}}$, where $t_{\text{cross-30}} = t_0 + 4338.6M$ is the 30th zero-crossing time of $\Re(h_{22})$. This transition interval spans about 28\% of the whole inspiral waveform.

\subsection{Line subtraction}
\label{sec:line_subtraction}

When we taper only the inspiral but not the ringdown, both ends of the waveform approach constants, but these two constants are generally different.  Because of the implicit assumption of periodicity in the DFT, this difference amounts to an unphysical discontinuity in the signal, which will pollute the spectrum with a component at just the right amplitude and phase dependence to interfere with the memory signal at low frequencies, and even swamp the memory at higher frequencies. Crucially, it does so in a way that depends sensitively on the arrival time of the GW signal, violating the standard assumptions in data analysis that allow us to optimize over arrival time using FFTs.  One simple option would be to also taper the end of the signal, but this would require a very gradual taper to avoid contaminating the low-frequency data with unphysical effects.  On the other hand, a gradual taper also implies that much of the signal is sacrificed to ``processing loss''~\cite{Harris:1978yaq}, which is the loss of signal power due to the tapering function reducing the signal amplitude toward zero near the boundaries.  This is a particular concern for the memory signal, which is already quite weak.  We can quite easily avoid these problems by simply subtracting the line connecting the two endpoints of the waveform.

To understand the problem posed by the discontinuity, we can construct a simple signal analogous to memory with a Heaviside step function, its value jumping from $0$ to some value $A$ at time $t=t_c$ representing the time of coalescence of the binary. This analogy is not new and is known as the zero-frequency limit used in, e.g., \cite{Smarr:1977fy, Turner:1978jj, Thorne:1992sdb}. If this signal is measured from $t=0$ to $t=T$, then the DFT will impose periodicity at these two instants and ``wrap'' around the signal from $t=T$ back to $t=0$.  The result is a ``pulse train'', rather than a Heaviside function, where the pulse is ``on'' for a time $T-t_c$.  Disregarding the irrelevant $f=0$ component, the Fourier transform of this signal is\footnote{Strictly speaking, this is true only in the limit of infinite sampling frequency.  For real signals, the true DFT will include aliasing that is most noticeable close to the Nyquist frequency (half the sampling frequency).}
\begin{align}
	\label{eq:pulse_fft}
	-A \frac{\sin (\pi f t_c) }{\pi f} e^{-\pi i f t_c}.
\end{align}
Note that this disagrees markedly with the ``shift theorem'' of Eq.~\eqref{eq:shift_theorem}: if we shift the signal in time (specifically, in the positive direction of the time axis) by some $\delta t$ while keeping the detection interval $ t\in [0, T] $ unchanged, then this is equivalent to a shift $ t_c \to t_c + \delta t $, and the phase of the Fourier transform changes by $-\pi f \delta t$, rather than the expected $-2\pi f \delta t$.  Even worse, the \emph{amplitude} of the transform explicitly depends on the time offset, which is not accounted for in the shift theorem.

We can resolve this seeming contradiction by noting that the shift theorem applies when the \emph{entire} signal is shifted by $\delta t$, which would have to include the discontinuity.  But this is not consistent with how the signal is actually constructed; the discontinuity remains at the same time (i.e., at $t=0$ and $t=T$), even when $t_c$ varies.  Therefore, the shift theorem does not apply in this case, which means that the standard matched-filtering approach of Eq.~\eqref{eq:matched_filter} is invalid for signals with memory.

Fortunately, we can recover the shift theorem with a simple modification to the preprocessing scheme: we take the line connecting the first and last points of the waveform, and subtract it from the signal.\footnote{There is a minor subtlety here. The DFT implicitly assumes periodicity in the signal and identifies the signal at times $t_0$ and $ t_{n-1} + \Delta t $ (not $ t_{n-1} $). Assuming the ringdown waveform has settled down to the value of $a_{n-1}$, the line to be subtracted is the one connecting the points $ (t_0, a_0) $ and $ (t_{n-1} + \Delta t, a_{n-1}) $.}  With this modification, the signal is now continuous---and in fact is smooth, assuming that derivatives of the original waveform were zero at the endpoints.  Its Fourier transform is\footnote{As with the expression in Eq.~\eqref{eq:pulse_fft}, this does not include aliasing effects.}
\begin{equation}
	\label{eq:sawtooth_fft}
	A \frac{e^{-2 \pi i f t_c}} {2 \pi i f}.
\end{equation}
Note that the amplitude is independent of the time offset, and the phase has the expected $-2\pi f \delta t$ dependence under the shift $ t_c \to t_c + \delta t $.  This is consistent with the shift theorem, so we can use the standard matched-filtering approach to search for the memory signal.

We will refer to this step as ``line subtraction'' to distinguish it from the very similar ``detrending'' that is found in the signal-analysis literature.\footnote{The \texttt{SciPy}~\cite{Virtanen:2019joe} and MATLAB~\cite{MATLAB_detrend_skip} packages, for example, implement functions named \texttt{detrend}.}  They must be distinguished because detrending fits a line (or occasionally some other polynomial) to the data, rather than connecting the first and last points.  Detrending is presumably much more sensible in noisy data, where the values \emph{precisely at} the beginning and end of the signal are essentially random.  However, applied to a model waveform with memory and no noise, detrending would still leave a discontinuity at the endpoints, and so would not be suitable for our purposes; we achieve better results with line subtraction.

Naturally, this simple step-function model is not a perfect representation of the true memory signal.  The smoother shape of the memory means that the high-frequency contents of the step function will not be present.  Indeed, in Sec.~\ref{sec:compare_other_procedures} we will examine both signals in the frequency domain, and see that they match at low frequencies, but the amplitude of the memory signal falls off at higher frequencies.  Nonetheless, it is still true that even the physical memory signal cannot be shifted in time arbitrarily and still be expected to match the physical data; if the shift is large enough that the memory (or any other component of a model waveform) changes at either the beginning or end of the data, then the shift theorem again does not apply.  This is the case even when dealing with models that do not include memory.

Also note that---as is standard~\cite{LIGOScientific:2019hgc, Allen:2005fk}---the same preprocessing procedure should be applied to the template waveform and signal data, in which case the line to be subtracted should be derived from the template rather than from the noisy data.  Now, the question arises as to what effect this subtracted line will have on the match [Eq.~\eqref{eq:matched_filter}] between the template and the data.  When maximizing $\langle s | h_{\delta t} \rangle$ over the time offset, line subtraction will not affect the \emph{location} of the optimum (the value of $\delta t$ that maximizes the inner product), because that quantity is maximized at the same location where the inner product of the difference $\langle s-h_{\delta t} | s-h_{\delta t} \rangle$ is minimized, which is invariant under simultaneous subtractions from $s$ and $h_{\delta t}$.  However, the \emph{value} of the match naturally will be affected, so it must be corrected before comparing to results for other template waveforms, after the difficult step of finding the optimal $\delta t$ has been done.  Fortunately, \emph{at the optimum} this correction is trivial to compute.  Let $A\cdot l(t)$ be the line that is subtracted in preprocessing, so that $l(t)$ is independent of the amplitude $A$. Then, the correction, which is the deviation of $\langle s-Al | h_{\delta t}-Al \rangle$ from $\langle s | h_{\delta t} \rangle$, consists of merely three integrals that need to be computed once per data stream per template, multiplied by $A$ or $A^2$.

\subsection{Standard procedure} \label{sec:standard_procedure}

We summarize our standard preprocessing procedure to be used in this paper as follows.
\begin{enumerate}
	\item Transition the last $20M$ of the time signal smoothly to a constant, using Eq.~\eqref{eqn:transition_to_constant}.
	\item Pad the right side by the end value for an additional time length of $20000M$.
	\item Taper the left side using the transition function Eq.~\eqref{eqn:transition_function} with $t_1 = t_0$ and $ t_2 = t_{\text{cross-30}}$.
	\item Subtract the signal by a line connecting its ends.
\end{enumerate}
Note that every step above is linear: Preprocessing a linear combination of time signals is the same as the linear combination of preprocessing component time signals separately. This linearity is very convenient when combined with the linearity of Fourier transform. For example, by Eq.~\eqref{eqn:mode_decomposition_plus}, $h_+(t)$ is the sum of $ H_{+,\ell m}(t) $. The linearity of preprocessing and Fourier transform guarantees that
\begin{align}
	\tilde{h}_+(f, \theta, \phi) = \sum_{\ell = 2}^{L} \sum_{|m|\le \ell} \tilde{H}_{+,\ell m}(f, \theta, \phi). \label{eqn:mode_decomposition_plus_freq}
\end{align}
This is the mode decomposition of the strain in the frequency domain, and will be quite useful in Sec.~\ref{sec:decomposition_spectrum}.

\subsection{Comparison with alternative procedures} \label{sec:compare_other_procedures}

We call our preprocessing procedure \textit{standard}, because we will apply it consistently throughout this paper. There are innumerable alternative ways to preprocess a waveform, and in this section, we illustrate the superiority of the standard procedure by comparing some of its variations. These variations include tapering with a bump window (Sec.~\ref{sec:tapering_bump_window}), using different padding lengths (Sec.~\ref{sec:vary_pad_length}), and using different tapering lengths (Sec.~\ref{sec:vary_taper_length}). In Sec.~\ref{sec:no_removal_initial_offsets}, we also compare the waveforms whose initial offsets are removed (see Sec.~\ref{sec:initial_offset_removal}) with those that still contain nonzero initial offsets.\footnote{We do not consider the step of initial offset removal as a part of the preprocessing procedure, because we need to carry out this step before evaluating the waveform along a direction $ (\theta, \phi) $. However, we still put the comparison between waveform spectra with and without initial offsets removed under Sec.~\ref{sec:compare_other_procedures}. The reason is that it is the tapering step that makes the removal step necessary.} 

For demonstration, we evaluate the waveform along the particular direction $ (\theta=\pi/2, \phi=0) $, i.e., edge-on view of the BBH. The $\times$ polarization of $h$ vanishes along this direction, so we only consider the + polarization $ h_+ $ and its components $ H_{+,\ell m} $. Note that the displacement memory is also the strongest along this direction.

\subsubsection{Tapering with a bump window} \label{sec:tapering_bump_window}

Alternative to the standard preprocessing scheme, we consider a scheme that tapers the waveform with a bump window Eq.~\eqref{eqn:bump_window} and has no line subtraction. We choose the same $t_1$ and $t_2$ from the inspiral tapering in the standard procedure, and choose $ t_4 = t_f + 20000M $ (the right end of the padded waveform) and $ t_3 = t_4-(t_2-t_1) $. For clarity, we refer to this new procedure as the \textit{bump window scheme}.

\begin{figure}[t]
	\centering
	\includegraphics[width=\linewidth]{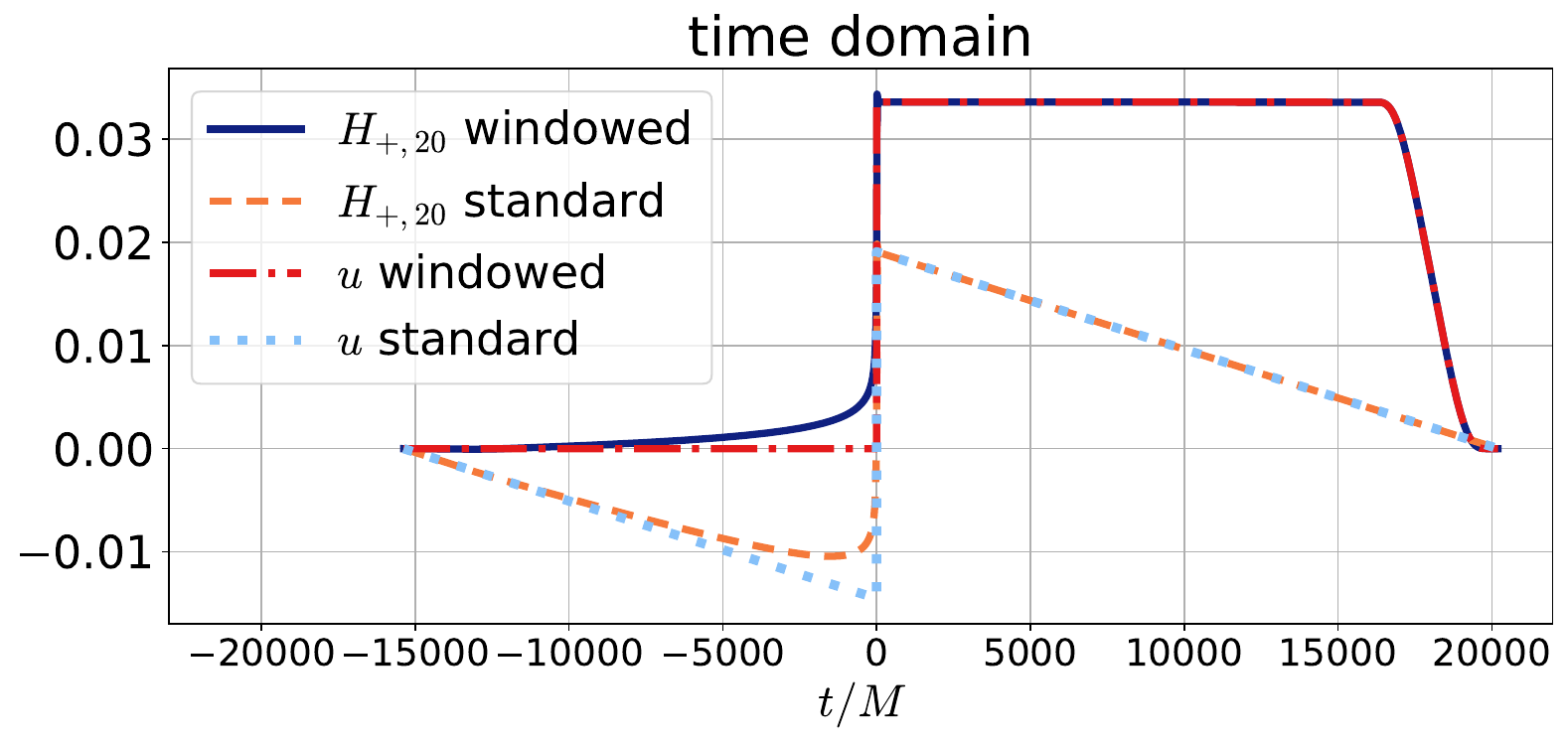}
	\includegraphics[width=\linewidth]{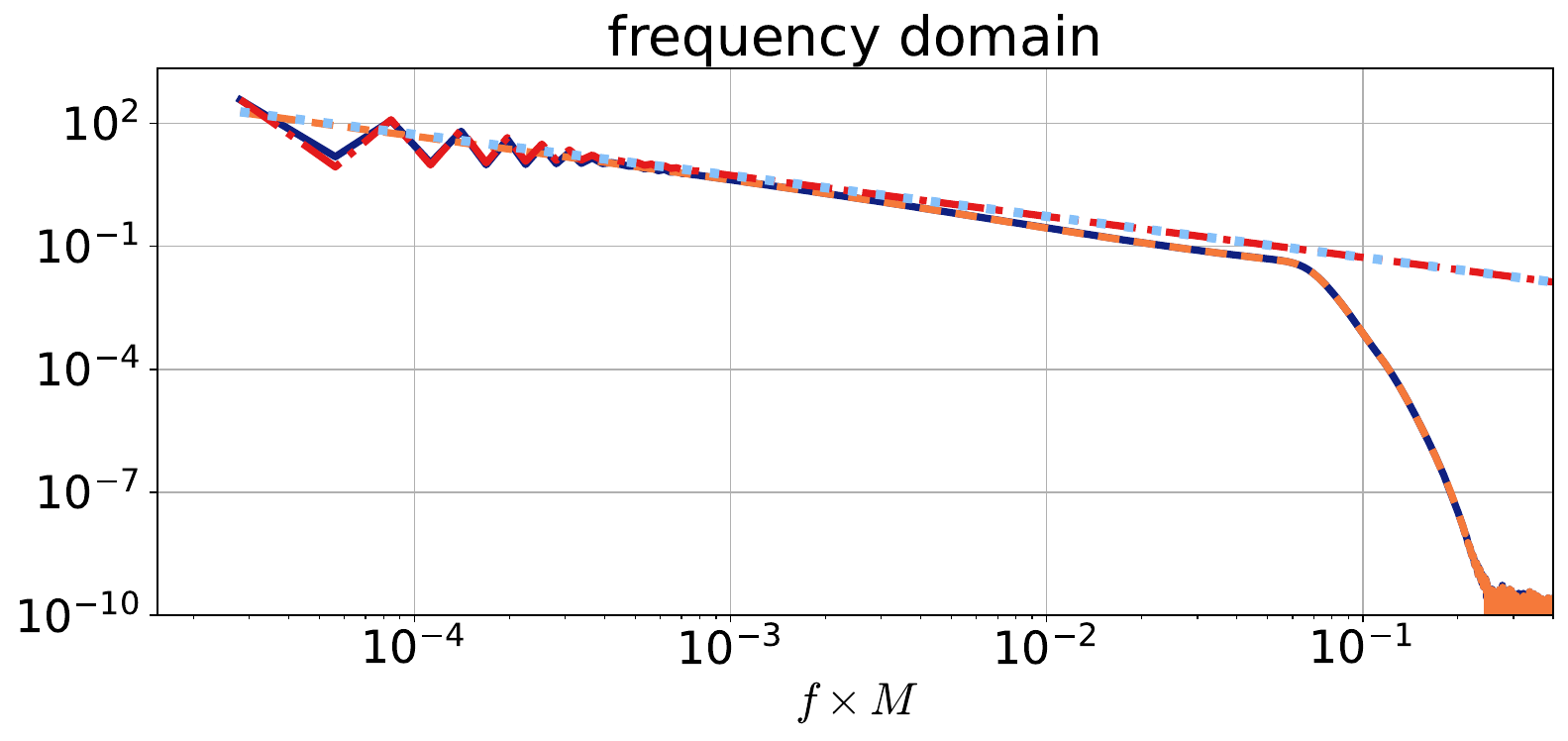}
	\caption{Comparison between the bump window scheme and the standard scheme, in the time domain (top) and frequency domain (bottom). The two schemes are introduced in Sec.~\ref{sec:tapering_bump_window}. The blue solid, orange dashed, red dash-dotted, and cyan dotted styles represent $ H_{+,20}(t) $ in the bump window scheme, $ H_{+,20}(t) $ in the standard scheme, $ u(t) $ in the bump window scheme, and $ u(t) $ in the standard scheme. The quantity $ u(t) $ is the Heaviside step function scaled by the right-end value of $ H_{+,20}(t) $.}
	\label{fig:preprocessing_line_subtract_or_not}
\end{figure}

We first consider the effects of these two schemes on $ H_{+, 20}(t, \theta=\pi/2, \phi=0) $ in the time domain. In the top panel of Fig.~\ref{fig:preprocessing_line_subtract_or_not}, the blue solid curve represents the bump window scheme, while the orange dashed curve represents the standard scheme. The blue curve resembles the original appearance of the non-preprocessed $H_{+, 20}$ (not shown), except that the right end is tapered off to zero. In contrast, the orange curve has a completely different looking due to line subtraction. However, this huge difference in the time domain is not carried over into the frequency domain. In fact, as shown in the bottom panel of Fig.~\ref{fig:preprocessing_line_subtract_or_not}, the two preprocessing schemes render the same frequency spectra after $ f = 5\times 10^{-4}/M $.\footnote{In the physical unit of frequency, $1/M \approx 203/(M/M_\odot)$~kHz, where $M_\odot$ is one solar mass. \label{fn:freq_conversion}} The only difference lies before $ f = 5\times 10^{-4}/M $, where the blue curve has zigzags while the orange curve remains straight.

We would like to investigate which scheme renders the spectrum more faithfully to the formally infinite-length waveform in the low frequency range. We can approximate the original time signal $H_{+, 20}$ as a Heaviside step function scaled by the end value of $H_{+, 20}$ (as we did in Sec.~\ref{sec:line_subtraction}). Let $u(t)$ be this scaled Heaviside step function. We preprocess $u(t)$ with the two schemes and show the preprocessed time signals in the top panel of Fig.~\ref{fig:preprocessing_line_subtract_or_not}, while the corresponding frequency spectra in the bottom panel. We use a red dash-dotted style to represent the bump window scheme, and a cyan dotted style for the standard scheme. In the frequency domain, we see that the red curve has a zigzag pattern similar to the blue curve, while the cyan curve remains straight. Because the spectrum of an infinitely long Heaviside step function is proportional to $ 1/f $ for all $f>0$ and specifically has neither zigzags nor oscillations, the standard scheme yields a more faithful spectrum in the low frequency range.

\subsubsection{Varying padding length} \label{sec:vary_pad_length}

When we pad a waveform by its end value, we introduce additional physical information into the signal. The longer we pad the waveform, the better the spectrum is resolved. However, limited by the computational power, we cannot pad a waveform arbitrarily longer. In this section, we investigate how the padding length affects the frequency spectrum.

\begin{figure}[t]
	\centering
	\includegraphics[width=\linewidth]{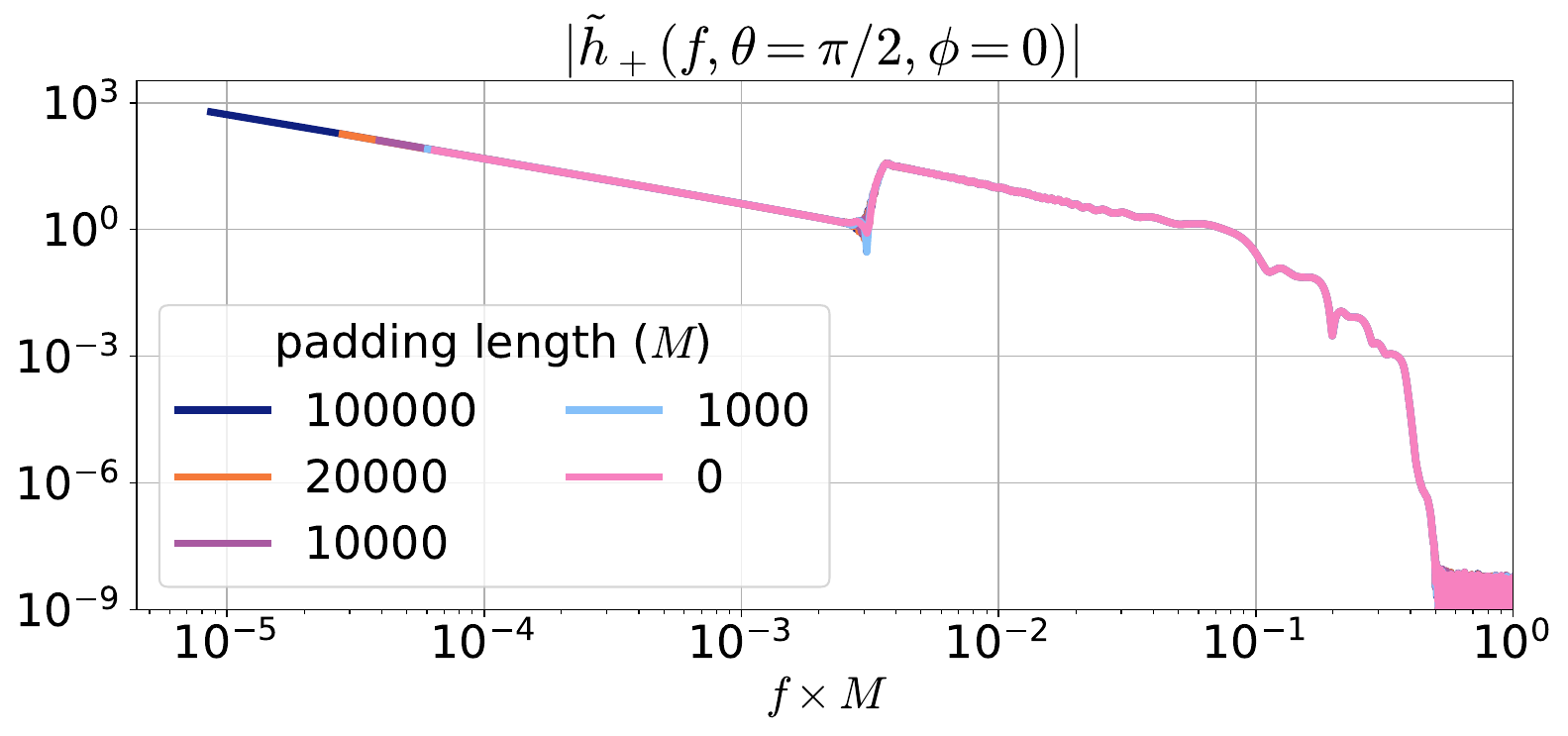}
	\includegraphics[width=\linewidth]{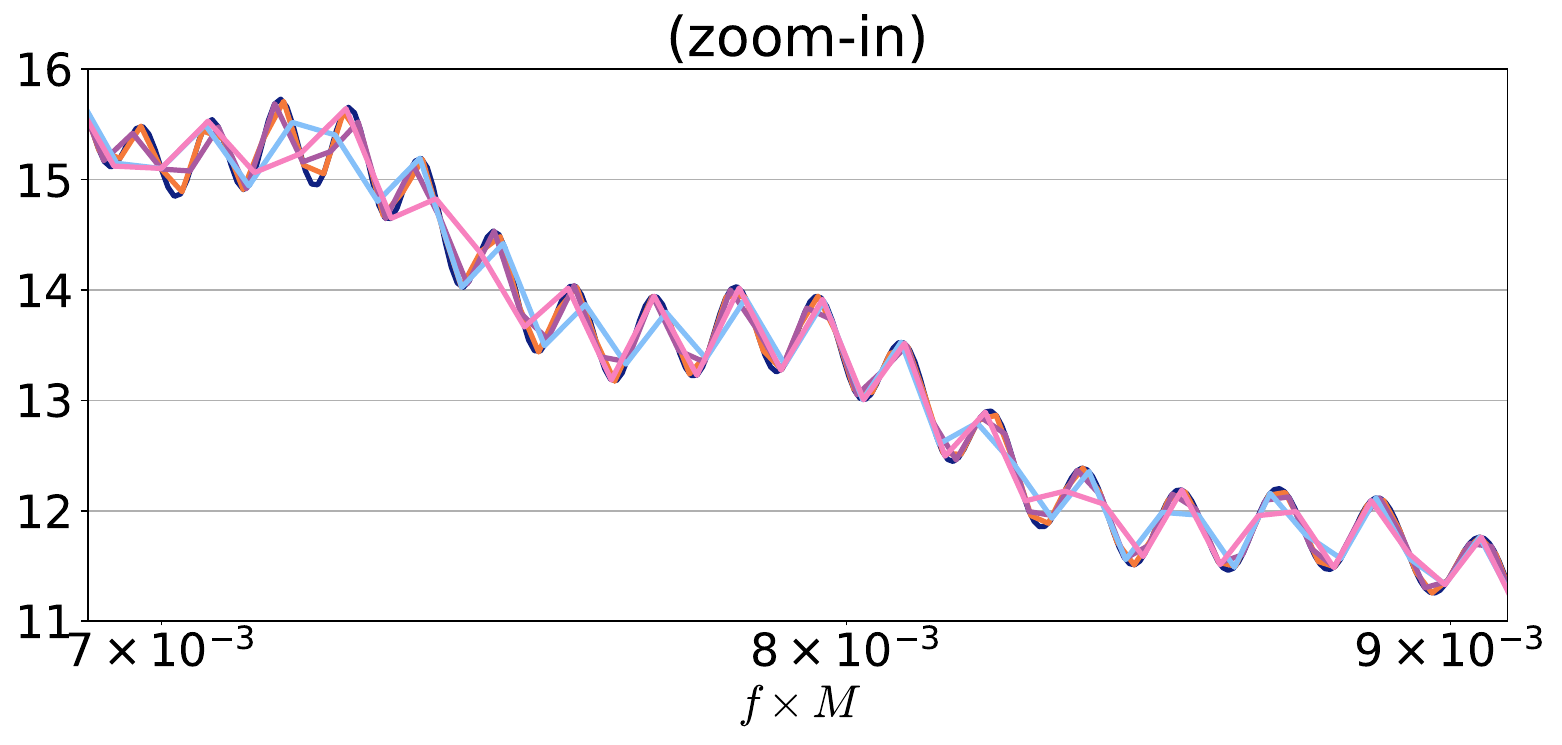}
	\caption{Frequency spectra of $ h_+(t, \theta=\pi/2, \phi=0) $ preprocessed with different padding lengths. The top panel shows the data at all positive frequencies, while the bottom panel is the close-up view of the top panel near $ f = 8\times 10^{-3}/M $. The five padding lengths are $100000M$ (blue), $20000M$ (orange), $10000M$ (purple), $1000M$ (cyan), and 0 (pink). We can see a jump in the spectrum near $ f = 3\times 10^{-3}/M $ in the top panel. We call it the ``turn-on'', and it originates from the finite length of the inspiral time signal.}
	\label{fig:preprocessing_vary_pad_length}
\end{figure}

We preprocess the signal $ h_+(t, \theta=\pi/2, \phi=0) $ using the standard procedure, except that we test five different padding lengths: 0, $1000M$, $10000M$, $20000M$, and $100000M$. Note that the padding length of 0 means no padding, and $20000M$ is our default padding length. We show the frequency spectra for different padding lengths in the top panel of Fig.~\ref{fig:preprocessing_vary_pad_length}. We find differences among these spectra in three regions. One is below $ f = 10^{-4}/M $, where we see that the spectrum with a longer padding length extends to lower frequencies. This happens because we only plot the data at positive frequencies and a longer time signal simply has a smaller frequency spacing in the spectrum. Another is the ``turn-on'' region near $ f = 3\times 10^{-3}/M $, where the inspiral spectrum starts to ramp up. This turn-on originates from the finite length of the inspiral time signal, so it is not physical but detection-wise inevitable. We can always move the turn-on region to a lower or higher frequency by generating a longer or shorter inspiral time signal. We can confirm that the tiny differences among the spectra in this region have the same origin as those differences in the third region we introduce in the next paragraph, so we do not show the zoom-in plot of the effect of various padding lengths on the turn-on region here.

The third region where we also find small differences among these spectra is in the inspiral part. These differences are hardly noticeable in the top panel of Fig.~\ref{fig:preprocessing_vary_pad_length}, so we zoom into a neighborhood of $ f=8\times10^{-3}/M $ and show it in the bottom panel. It is obvious that the five curves on this graph are merely five different samples of a common underlying continuous curve. A longer padding yields a finer sampling in the frequency domain, which helps us better identify any potential features in the spectrum. The curves representing the padding lengths of $0M$ (pink) and $1000M$ (cyan) contain so many zigzags that we may overlook the oscillatory physical features and treat them as nonphysical Gibbs behaviors. The $100000M$-padding-length curve (blue) has the best resolution, but it may overkill and compromise computational speed. Therefore, we believe that a padding length of $20000M$ produces a sufficiently resolved spectrum for this paper.

\subsubsection{Varying tapering length} \label{sec:vary_taper_length}

In the standard preprocessing procedure, the tapering roll-on starts from the left end of the time signal and ends at the 30th zero-crossing of $ \Re (h_{22}) $. The tapering length is $ t_2- t_1 = 4338.6M $. In this section, we investigate how a different tapering length can affect the frequency spectrum of the waveform $ h_+(t, \theta=\pi/2, \phi=0) $. We fix $t_1 = t_0$ (the left end point) and vary $t_2$. 

\begin{figure}[t]
	\centering 
	\includegraphics[width=\linewidth]{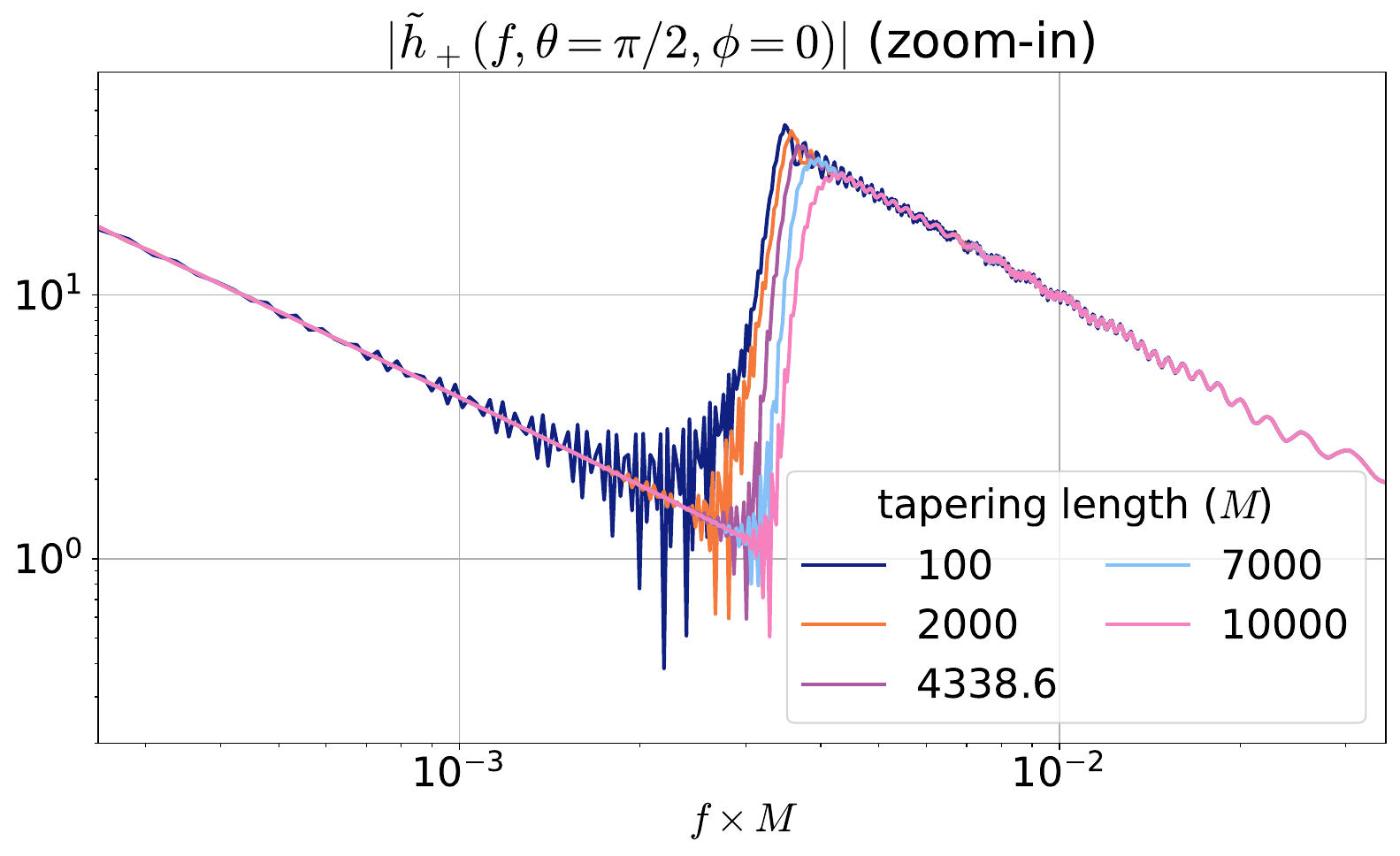}
	\caption{Frequency spectra of $ h_+(t, \theta=\pi/2, \phi=0) $ preprocessed with five different tapering lengths: $100M$ (blue), $2000M$ (orange), $4338.6M$ (purple), $7000M$ (cyan), and $10000M$ (pink). Note that $4338.6M$ is the default value used in our standard procedure. This graph only shows the turn-on region.}
	\label{fig:preprocessing_vary_taper_rollon}
\end{figure}

We choose five different tapering lengths: $100M$, $2000M$, $4338.6M$, $7000M$, and $10000M$. We find that the frequency spectra only differ near the turn-on region, as shown in Fig.~\ref{fig:preprocessing_vary_taper_rollon}. Because the turn-on region can fall in the sensitivity band of a detector, we would like to confine any spurious effects in this region as narrowly as possible. To achieve this goal, we can apply longer tapering whose spectral leakage is milder. Indeed, we see from the graph that a longer tapering leads to fewer oscillations in the turn-on region. However, we can also see the side effect of using a longer tapering: Because more inspiral contents in the time domain are tapered away, the inspiral spectrum spans a narrower frequency range. We consider the tapering length $4338.6M$ (purple) a good choice, for it produces little spectral leakage near the turn-on region without eroding much inspiral spectrum.

\subsubsection{No removal of initial offsets} \label{sec:no_removal_initial_offsets}

\begin{figure}[t]
	\centering 
	\includegraphics[width=\linewidth]{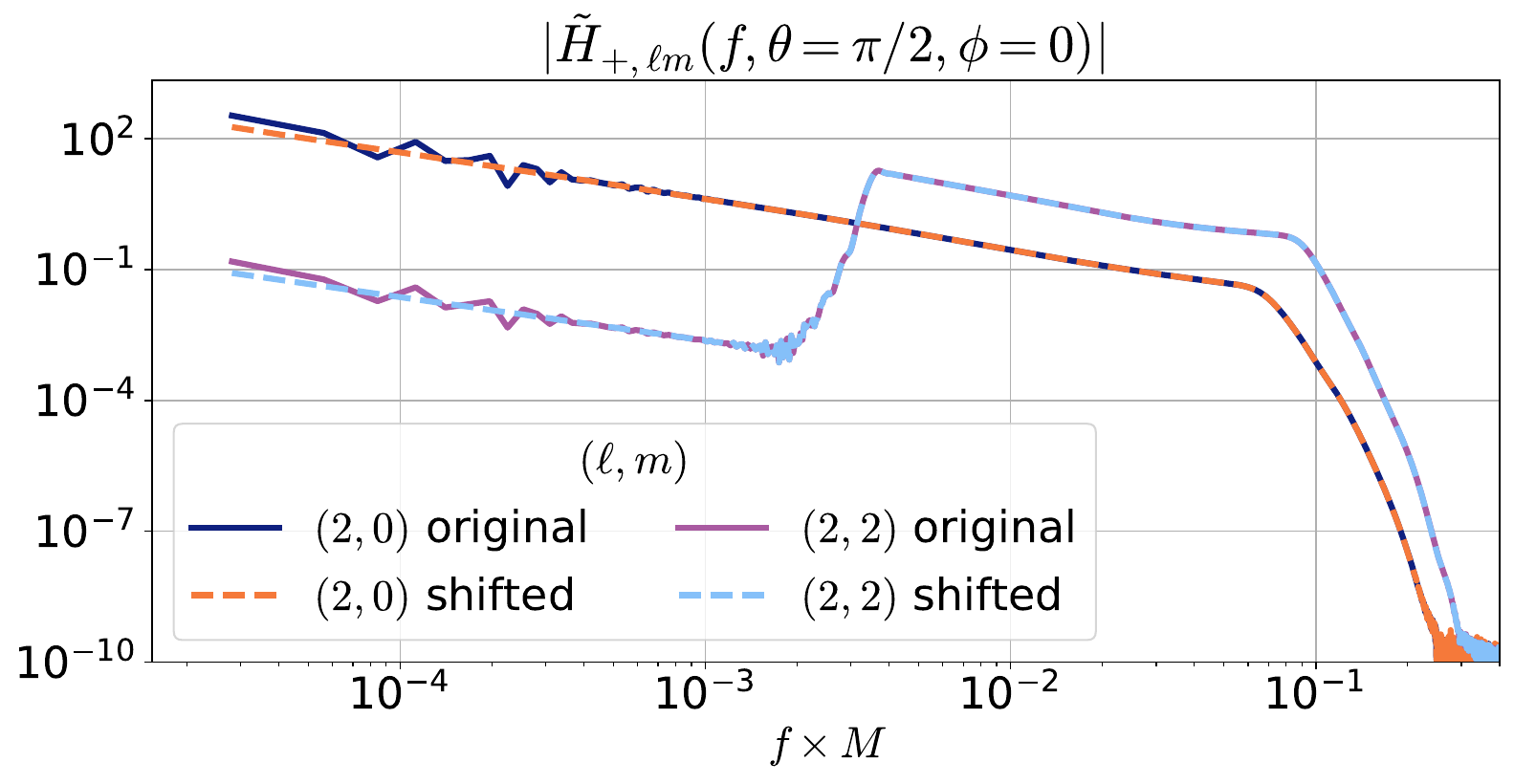}
	\caption{Frequency spectra of two waveform components with and without initial offsets being removed. Both the spectra whose initial offsets are not removed (blue solid and purple solid) have zigzag patterns in the low frequency range. In contrast, the spectra with initial offsets being removed (orange dashed and cyan dashed) remain straight in this range.}
	\label{fig:preprocessing_shift_J_E_time0_values}
\end{figure}

As described in Sec.~\ref{sec:initial_offset_removal}, we remove constant values from waveform modes so that the electric null parts of all waveform modes start at 0, i.e., $ J_{\mathcal{E}, \ell m}(t=t_0) = 0 $. We demonstrate the necessity of this removal step in Fig.~\ref{fig:preprocessing_shift_J_E_time0_values} using two concrete examples: the spectra of $ H_{+,20} $ and $ H_{+,22} $. The spectrum of the original $ H_{+,20} $ (whose initial offset is not removed, blue solid) is different from the spectrum of the shifted $ H_{+,20} $ (whose initial offset is removed, orange dashed) in the low frequency range.\footnote{Note that the original $ H_{+,20} $ and $ H_{+,22} $ settle down to 0 at late times, while the shifted ones do not. Figure~\ref{fig:memory_modes} plots the shifted ones.} Specifically, the blue curve has a zigzag pattern while the orange curve remains straight. Because the low frequency part of $ |\tilde{H}_{+,20}| $ should approach the zero-frequency limit $\propto 1/f $ (see Sec.~\ref{sec:tapering_bump_window}), removing the initial offset indeed improves the frequency spectrum of $ H_{+,20} $. We also see a very similar behavior in the spectra of $ H_{+,22} $ (purple solid and cyan dashed). This is in consistence with the presence of the memory offset in $h_{22}$ (see the bottom right panel of Fig.~\ref{fig:memory_modes}).

\section{Mode decomposition of a waveform spectrum} \label{sec:decomposition_spectrum}

In Sec.~\ref{sec:preprocessing}, we carefully choose the appropriate preprocessing procedure that reduces spurious contents in the frequency spectrum of a waveform. As an application of our standard preprocessing scheme, we shift our attention to the spectrum itself and search for any interesting features therein. For simplicity, we only consider the direction $ (\theta=\pi/2, \phi=0) $ so the strain is  + polarized.

\begin{figure}[t]
	\centering
	\includegraphics[width=\linewidth]{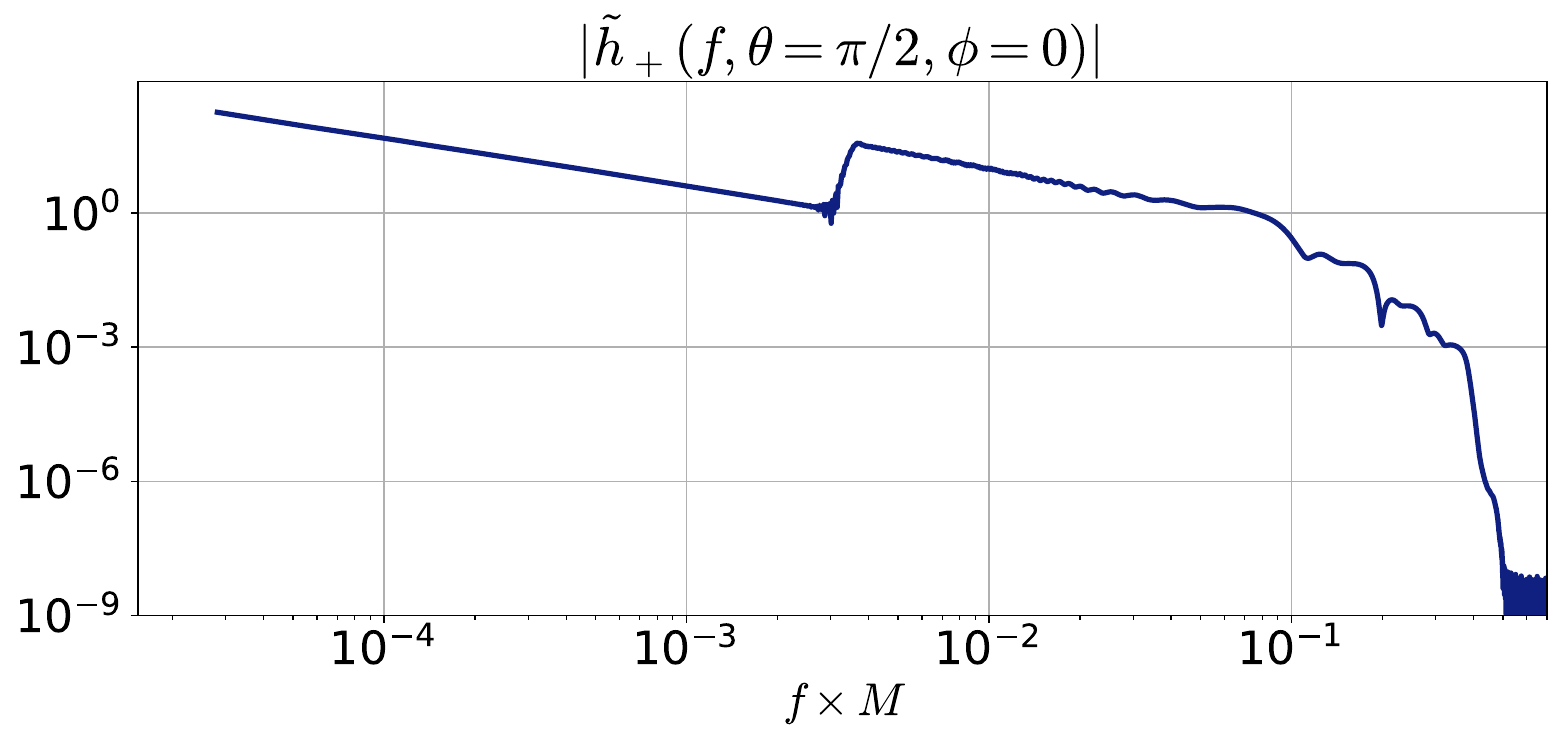}
	\includegraphics[width=\linewidth]{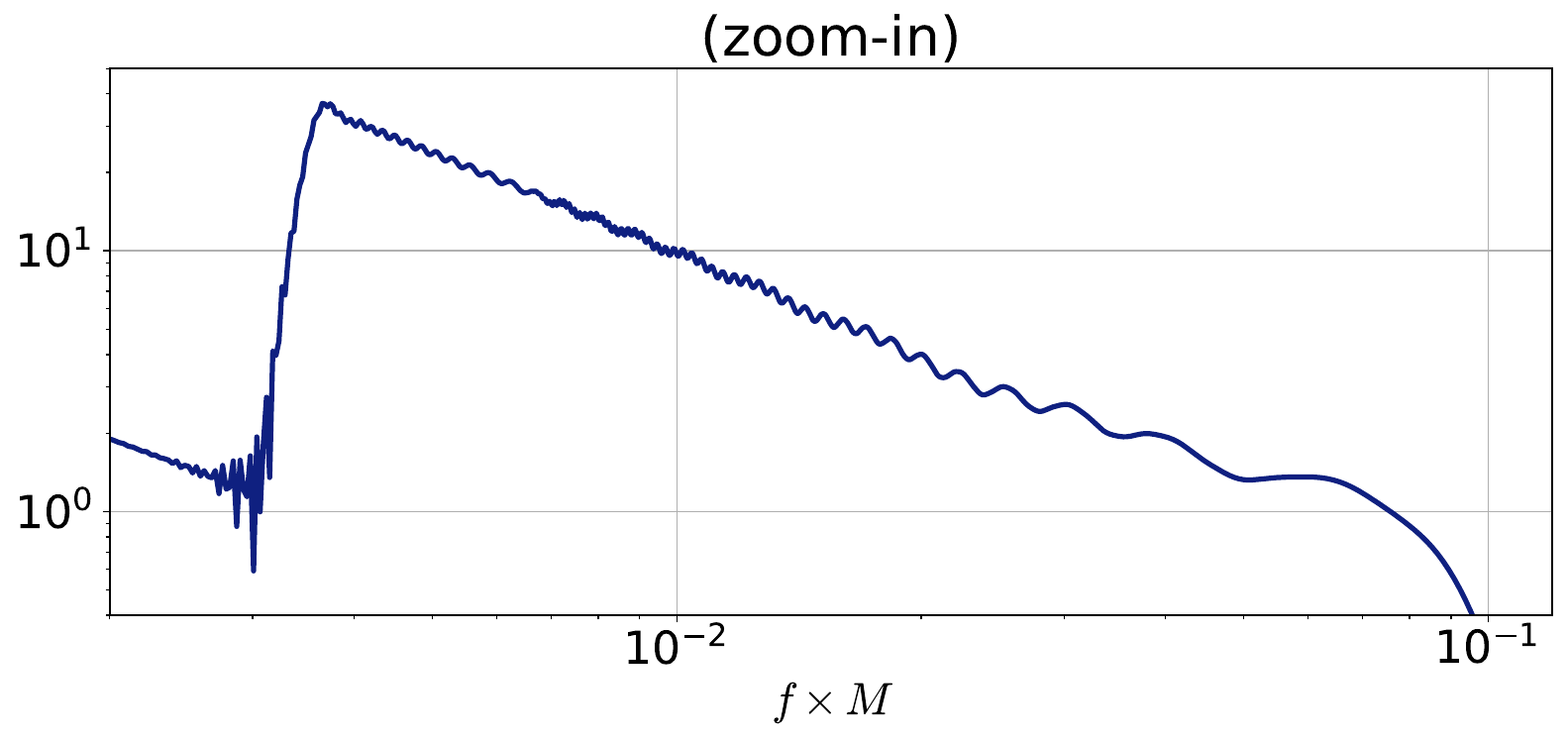}
	\caption{Frequency spectrum of $ h_+(f, \theta=\pi/2, \phi=0) $, preprocessed using the standard procedure in Sec.~\ref{sec:standard_procedure}. The top panel shows the full range $f>0$, while the bottom panel zooms into the inspiral part.}
	\label{fig:decomposition_overall}
\end{figure}

\begin{figure*}[t]
	\centering
	\includegraphics[width=0.496\linewidth]{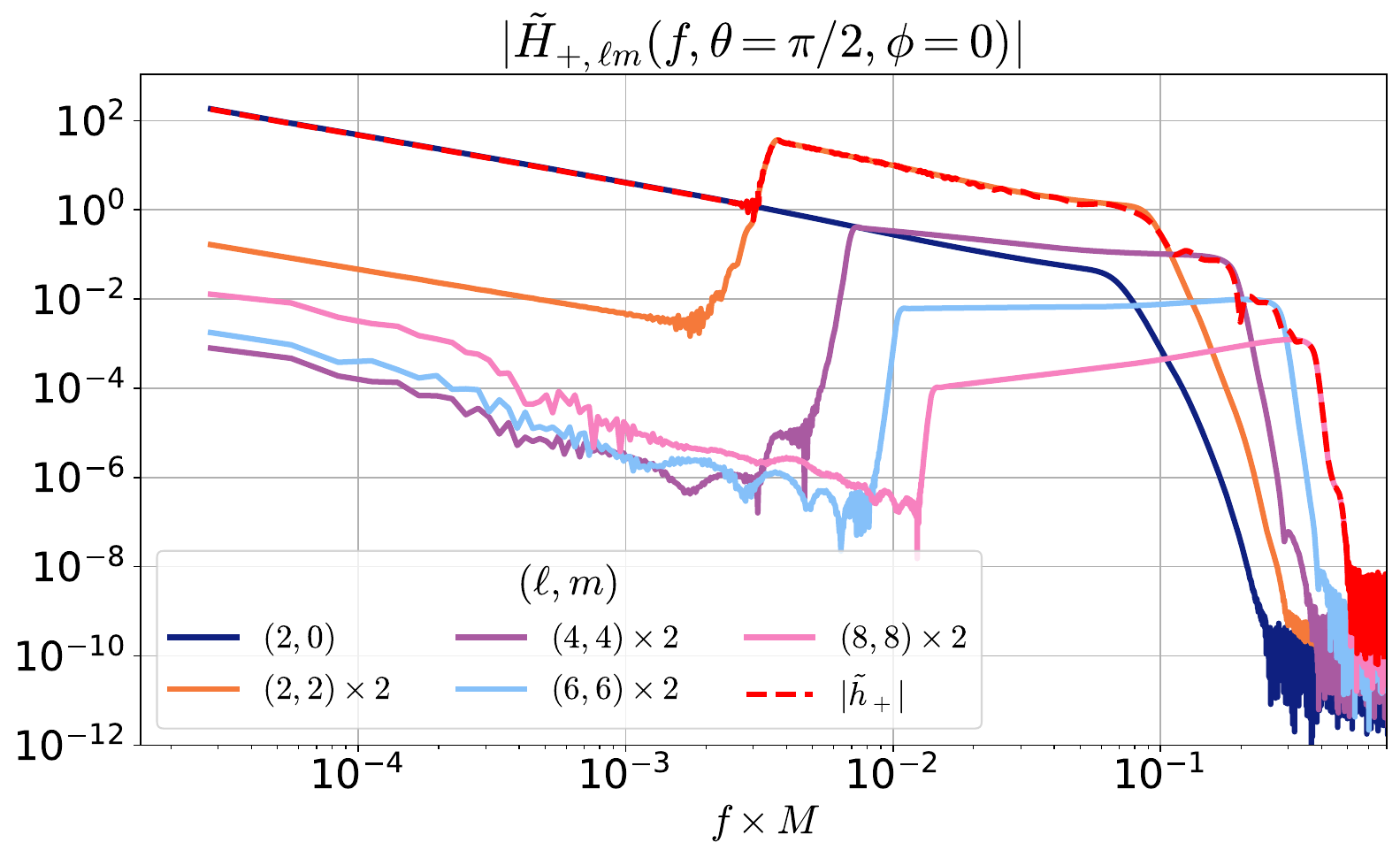}
	\includegraphics[width=0.496\linewidth]{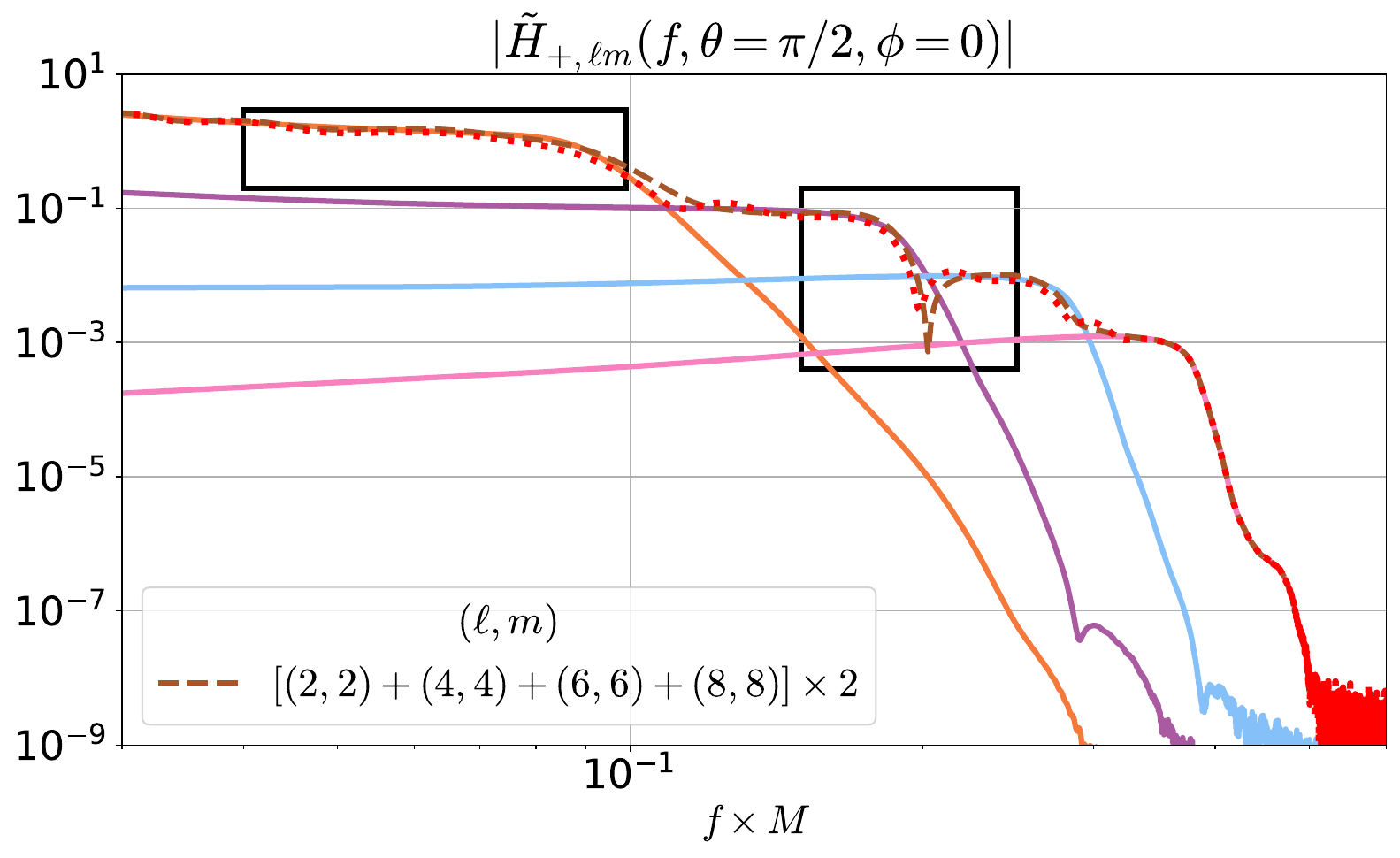}
	\caption{Left: main contributions of waveform components to the overall spectrum. Right: enlarged view of the ringdown spectrum. There are seven curves in total: $ |\tilde{h}_+| $ (red dashed), $ |\tilde{H}_{+,20}| $ (blue), $ |\tilde{H}_{+,22} + \tilde{H}_{+,2(-2)}| $ (orange), $ |\tilde{H}_{+,44} + \tilde{H}_{+,4(-4)}| $ (purple), $ |\tilde{H}_{+,66} + \tilde{H}_{+,6(-6)}| $ (cyan), $ |\tilde{H}_{+,88} + \tilde{H}_{+,8(-8)}| $ (pink), and $ |\tilde{H}_{+,22} + \tilde{H}_{+,2(-2)} + \tilde{H}_{+,44} + \tilde{H}_{+,4(-4)} + \tilde{H}_{+,66} + \tilde{H}_{+,6(-6)} + \tilde{H}_{+,88} + \tilde{H}_{+,8(-8)}| $ (brown dashed). The symbol ``$\times 2$'' in the legend is the scaling of the quantities. For example, the label $ (2,2) \times 2 $ refers to the quantity $ |\tilde{H}_{+,22} + \tilde{H}_{+,2(-2)}| = 2|\tilde{H}_{+,22}| $. The right panel also contains two boxed regions that are introduced in Sec.~\ref{sec:contribution_components}.}
	\label{fig:decomposition_shapes}
\end{figure*}

The top panel of Fig.~\ref{fig:decomposition_overall} shows the frequency spectrum of $ h_+(t, \theta=\pi/2, \phi=0) $. This curve is by construction identical to the $20000M$ curve (orange) in the top panel of Fig.~\ref{fig:preprocessing_vary_pad_length}. We can split the spectrum into three segments: the $1/f$ part ($ f < 3\times 10^{-3}/M $), the inspiral part ($ 3\times 10^{-3}/M < f < 0.06/M $), and the ringdown part ($ f> 0.06/M $). In Sec.~\ref{sec:contribution_components}, we will see how these segments are shaped by only a handful of $ (\ell, m) $ components. We also show the enlarged view of the inspiral part in the bottom panel of Fig.~\ref{fig:decomposition_overall}. This part contains many small oscillations, which the readers may have noticed earlier in Figs.~\ref{fig:preprocessing_vary_pad_length} and \ref{fig:preprocessing_vary_taper_rollon}. We will investigate the origin of these oscillations in Sec.~\ref{sec:beats}.

A powerful tool used in this section is the mode decomposition of the DFT of the waveform, Eq.~\eqref{eqn:mode_decomposition_plus_freq}. This means that the frequency signals of the waveform components, $ \tilde{H}_{+,\ell m} $, are additive, and they sum up to the overall frequency signal, $ \tilde{h}_+ $. Although the magnitudes $ |\tilde{H}_{+,\ell m}| $ do not sum up to $ |\tilde{h}_+| $ and we do mostly care about the magnitudes (e.g., $ |\tilde{h}_+| $) rather than the frequency signal itself (e.g., $ \tilde{h}_+ $), Eq.~\eqref{eqn:mode_decomposition_plus_freq} is still very instructive, especially when a component dominates the others.

\subsection{Contributions to a waveform spectrum} \label{sec:contribution_components}

The overall spectrum $ | \tilde{h}_+ | $ is mainly determined by nine waveform components, as shown in the left panel of Fig.~\ref{fig:decomposition_shapes}. This graph contains a copy of $ | \tilde{h}_+ | $ (red dashed) and another five curves (solid). For each $ m > 0 $ component, we plot $ |\tilde{H}_{+,\ell m} + \tilde{H}_{+,\ell(-m)}|$ rather than $ |\tilde{H}_{+,\ell m}| $, because for our BBH configuration, both the $ (\ell, m) $ and $ (\ell,-m) $ components have identical contributions to the overall spectrum, i.e.,
\begin{align}
	H_{+,\ell m}(t,\theta=\pi/2,\phi=0) = H_{+,\ell (-m)}(t,\theta=\pi/2,\phi=0) \label{eqn:hlm_equal_hl-m}.
\end{align}
This explains the labels ``$ (\ell, m) \times 2 $'' in the legend of the graph. For example, the orange curve is the spectrum $ |\tilde{H}_{+,22}|\times 2 $, representing the contribution from the $ (2, \pm 2) $ components. As every $ m > 0 $ curve in the graph contains contribution of two components, all five solid curves together represent nine components.

Different waveform components dominate different segments of $ | \tilde{h}_+ | $. In the left panel of Fig.~\ref{fig:decomposition_shapes}, we find that the $ (2,0) $ component (blue) dominates the $1/f$ part of $ | \tilde{h}_+ | $ (red dashed). Because $ \tilde{H}_{+,20} $ approaches $ 1/f $ in the low frequency range (see Sec.~\ref{sec:tapering_bump_window}), so should the overall spectrum, which justifies the name ``$ 1/f $ part''. We also find that the $ (2,\pm 2) $ components (orange) dominate the inspiral part of $ | \tilde{h}_+ | $, as expected.

The ringdown part is more complicated, as the dominant contribution includes eight components: $ (2,\pm2) $, $ (4,\pm4) $, $ (6,\pm6) $, and $ (8,\pm8) $. There could be contribution from higher $ \ell = m $ components in theory, but the numerical waveform in this paper only contains $ \ell \le 8 $. We provide a close-up view of the ringdown part in the right panel of Fig.~\ref{fig:decomposition_shapes}. We can see from the graph that, as $m$ increases, these components extend to a higher-frequency range but with a lower magnitude. The resulting 8-component combined spectrum (brown dashed) is thus a stepped decreasing function. This 8-component spectrum deviates slightly from the overall spectrum (red dashed), because of the presence of other components in the overall spectrum. We checked that the contribution of the $ (3,\pm2) $ components is the main reason why the overall spectrum is slightly lower than the 8-component spectrum in the range $ 0.04/M \le f \le 0.1/M $ (enclosed in the left black box). We also checked that the deviation between these two curves (the 8-component spectrum and the overall spectrum) in the range $ 0.15/M \le f \le 0.25/M $ (enclosed in the right black box) can be attributed to the $ (5, \pm4) $ components.

\begin{figure*}[t]
	\centering
	\includegraphics[width=0.496\linewidth]{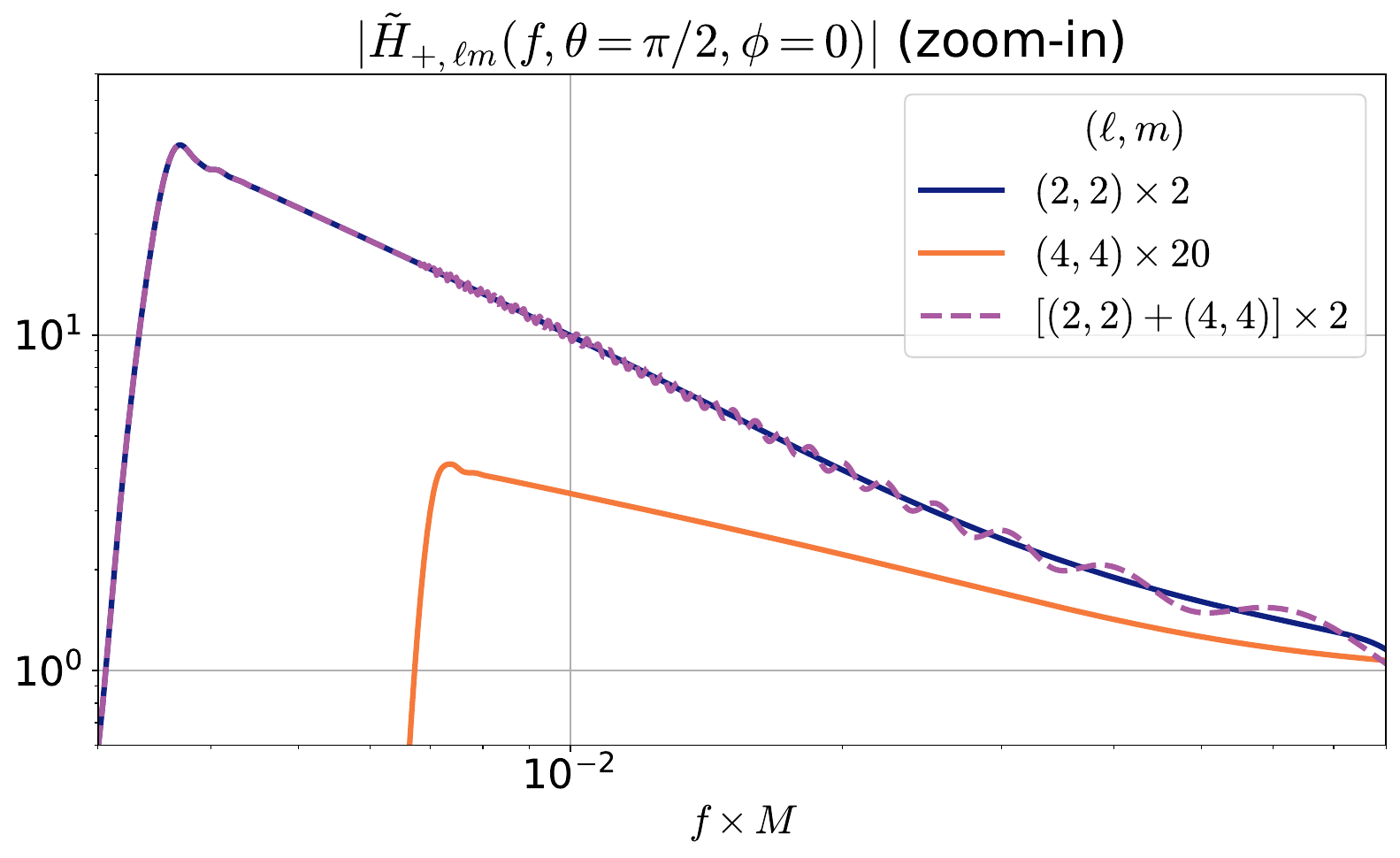}
	\qquad
	\includegraphics[width=0.34\linewidth]{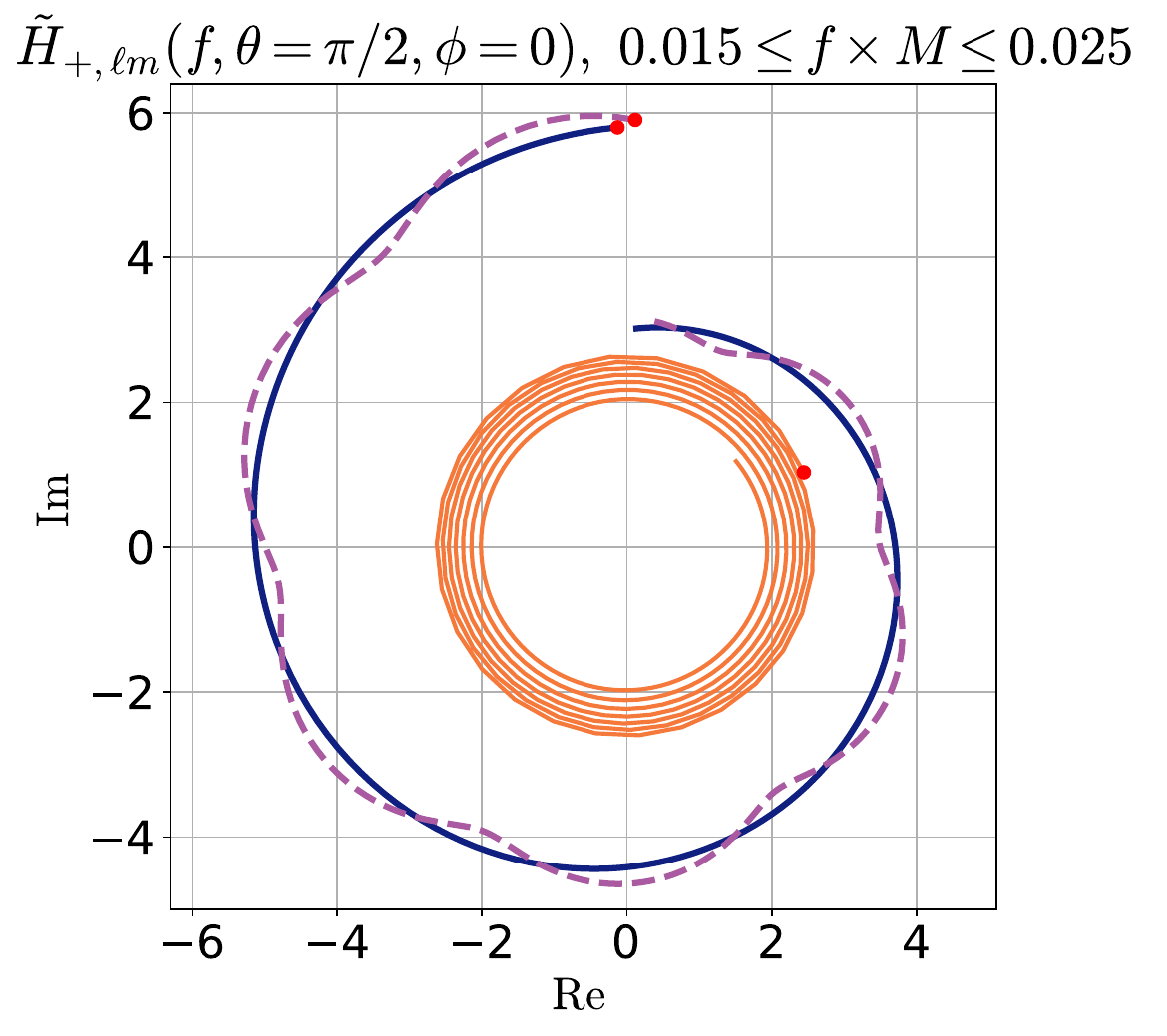}
	\includegraphics[width=0.496\linewidth]{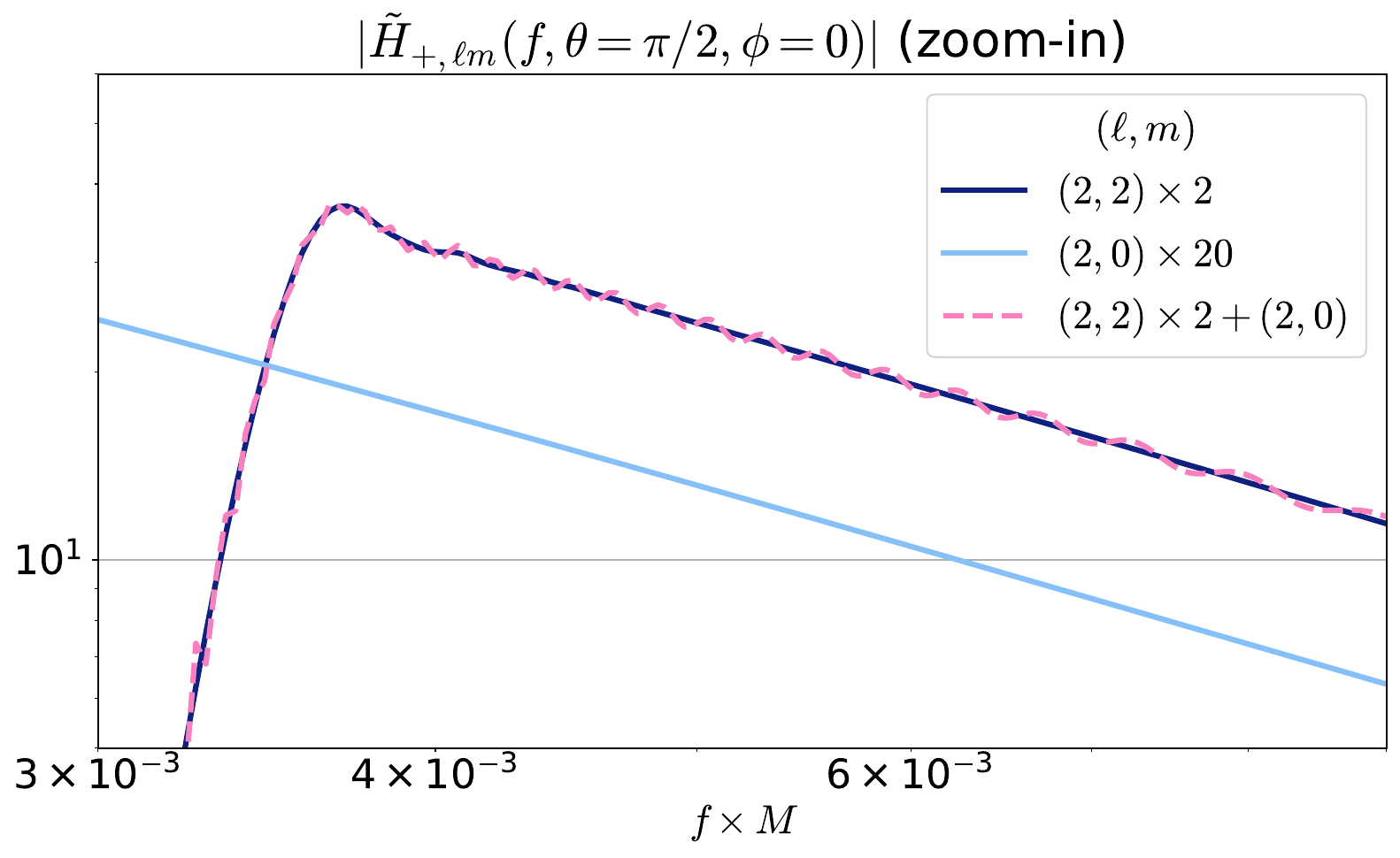}
	\qquad
	\includegraphics[width=0.34\linewidth]{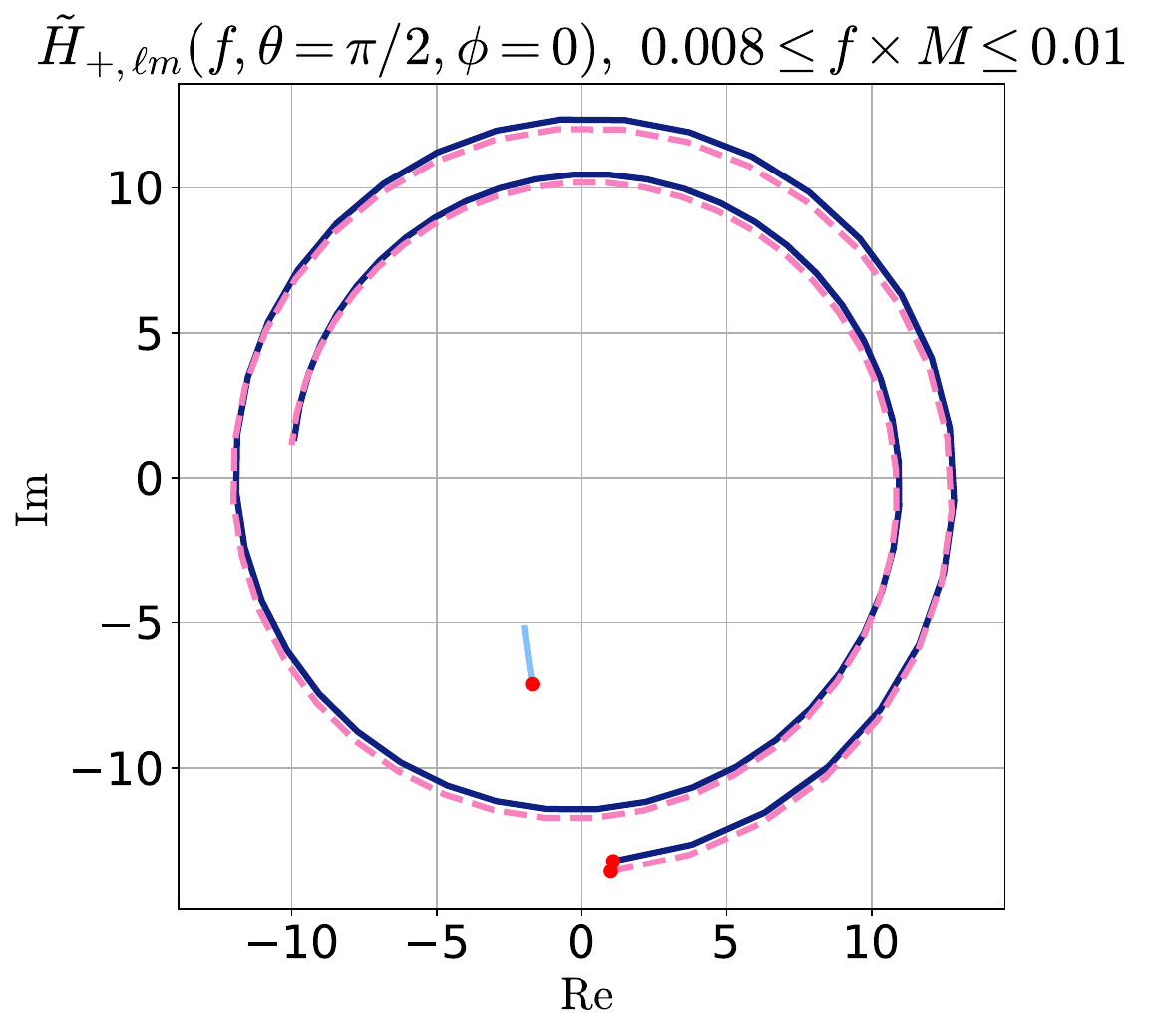}
	\caption{Beats formed by the $ (2,\pm 2) $, $ (4,\pm 4) $, and $ (2,0) $ waveform components. The right column shows their frequency signals on a complex plane, while the left column shows their amplitude spectra. These frequency signals are $ \tilde{H}_{+,22} + \tilde{H}_{+,2(-2)} $ (blue), $ 10[\tilde{H}_{+,44} + \tilde{H}_{+,4(-4)}] $ (orange), $ \tilde{H}_{+,22} + \tilde{H}_{+,2(-2)} + \tilde{H}_{+,44} + \tilde{H}_{+,4(-4)} $ (purple dashed), $ 20\tilde{H}_{+,20}$ (cyan), and $ \tilde{H}_{+,22} + \tilde{H}_{+,2(-2)} + \tilde{H}_{+,20} $ (pink dashed). The red dots in the top right panel stand for the data at $ f=0.015/M $, and the red dots in the bottom right panel stand for $ f=0.008/M $. The symbols ``$\times 2$'' and ``$\times 20$'' in the legend stand for the scaling of the corresponding quantities. For example, the label $ (2,2) \times 2 $ refers to the quantity $ \tilde{H}_{+,22} + \tilde{H}_{+,2(-2)} = 2\tilde{H}_{+,22} $, and the label $ (4,4) \times 20 $ refers to the quantity $ 10[\tilde{H}_{+,44} + \tilde{H}_{+,4(-4)}] = 20\tilde{H}_{+,44} $.}
	\label{fig:decomposition_beats}
\end{figure*}

\subsection{Oscillations in the inspiral spectrum} \label{sec:beats}

The small oscillations in the inspiral part of the overall spectrum (see the bottom panel of Fig.~\ref{fig:decomposition_overall}) are the results of two beating patterns. One beat is formed by the $ (2,\pm 2) $ and $ (4,\pm 4) $ components, while the other beat by the $ (2,\pm 2) $ and $ (2,0) $ components.

To inspect the first beat, we show the spectra of the relevant components in the top left panel of Fig.~\ref{fig:decomposition_beats}. The blue curve represents the spectrum of $ H_{+,22} + H_{+,2(-2)} $, which is labeled as ``$ (2,2)\times 2 $'' for the same reason explained in the text beneath Eq.~\eqref{eqn:hlm_equal_hl-m}. The orange curve is the spectrum of $ H_{+,44} + H_{+,4(-4)} $, which is scaled up by a factor of 10 so that we can fit the blue and orange curves in the same graph without losing details of their combined spectrum (purple dashed). We see that both the blue and orange curves have no oscillations, but their combination does oscillate, starting from the turn-on frequency of the $ (4,\pm4) $ components.

The top right panel shows the frequency signals on a complex plane,\footnote{In Sec.~\ref{sec:bbh_simulation}, we have shifted the time axis so that $t=0$ corresponds to the merger time instead of the initial time. According to the shift theorem Eq.~\eqref{eq:shift_theorem}, this time shift incurs an $f$-dependent phase factor in the DFT of the strain. We do take this phase factor into account. A time shift does not affect the amplitude spectrum, but can change the appearance of the frequency signal on a complex plane.} for $ 0.015/M \le f \le 0.025/M $. The amplitude of the $ (4,\pm 4) $-component curve (orange) is scaled up by a factor of 10, consistent with the top left panel. As frequency increases, both the blue and orange curves spiral in smoothly, but the sum of them, the purple dashed curve, crosses the blue curve regularly. This behavior is translated to a beat pattern in the amplitude spectrum. 

We inspect the second beat in the same manner. The bottom left panel of Fig.~\ref{fig:decomposition_beats} shows the spectra of $ H_{+,22} + H_{+,2(-2)} $ (blue), $ H_{+,20} $ (cyan, scaled up by a factor of 20), and their combination (pink dashed). The bottom right panel shows the corresponding frequency signals on a complex plane for $ 0.008/M \le f \le 0.01/M $. We see from the bottom right panel that the $ (2,0) $ component is approximately a constant in this frequency range. Adding this constant to the blue curve generates a beating effect in the amplitude spectrum.

\section{Detecting gravitational-wave memory} \label{sec:detecting_gw}

As another application of using line subtraction instead of ringdown tapering in preprocessing time signals, we investigate the detection of the memory component of GWs. In Sec.~\ref{sec:snr_ligo_ce_results}, we calculate the SNRs of simulated GW signals with and without memory in O4's LIGO and CE. In Sec.~\ref{sec:snr_lisa_results}, we calculate the SNRs of the signals in LISA, and propose a simple technique to detect memory in a LISA signal in a special scenario. In Sec.~\ref{sec:result_param_infer}, we use Bayesian inference to analyze possible scenarios of memory detection in LIGO and CE.

\subsection{Signal-to-noise ratios in LIGO and CE} \label{sec:snr_ligo_ce_results}

\begin{figure*}[t]
	\centering
	\includegraphics[width=0.496\linewidth]{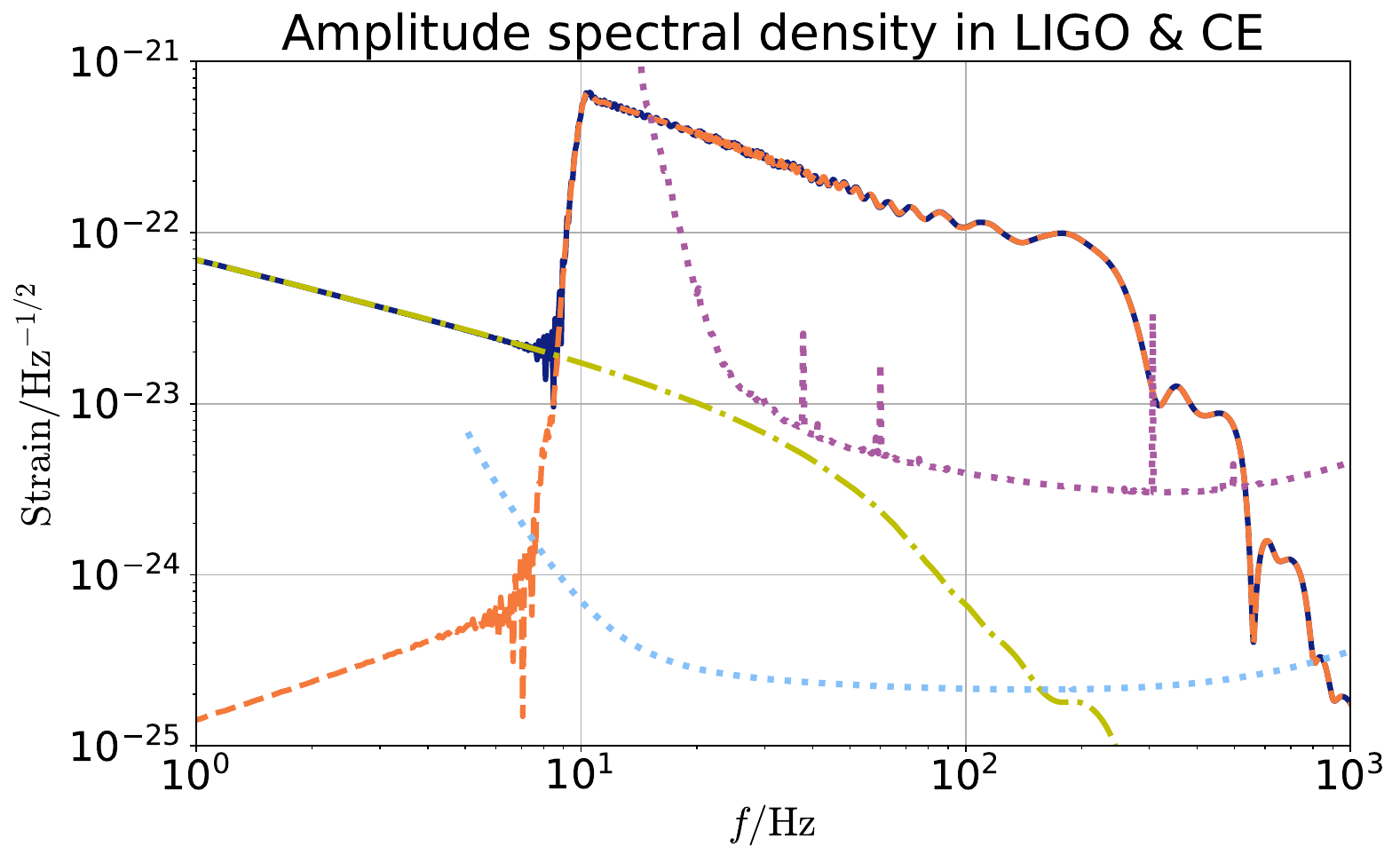}
	\includegraphics[width=0.496\linewidth]{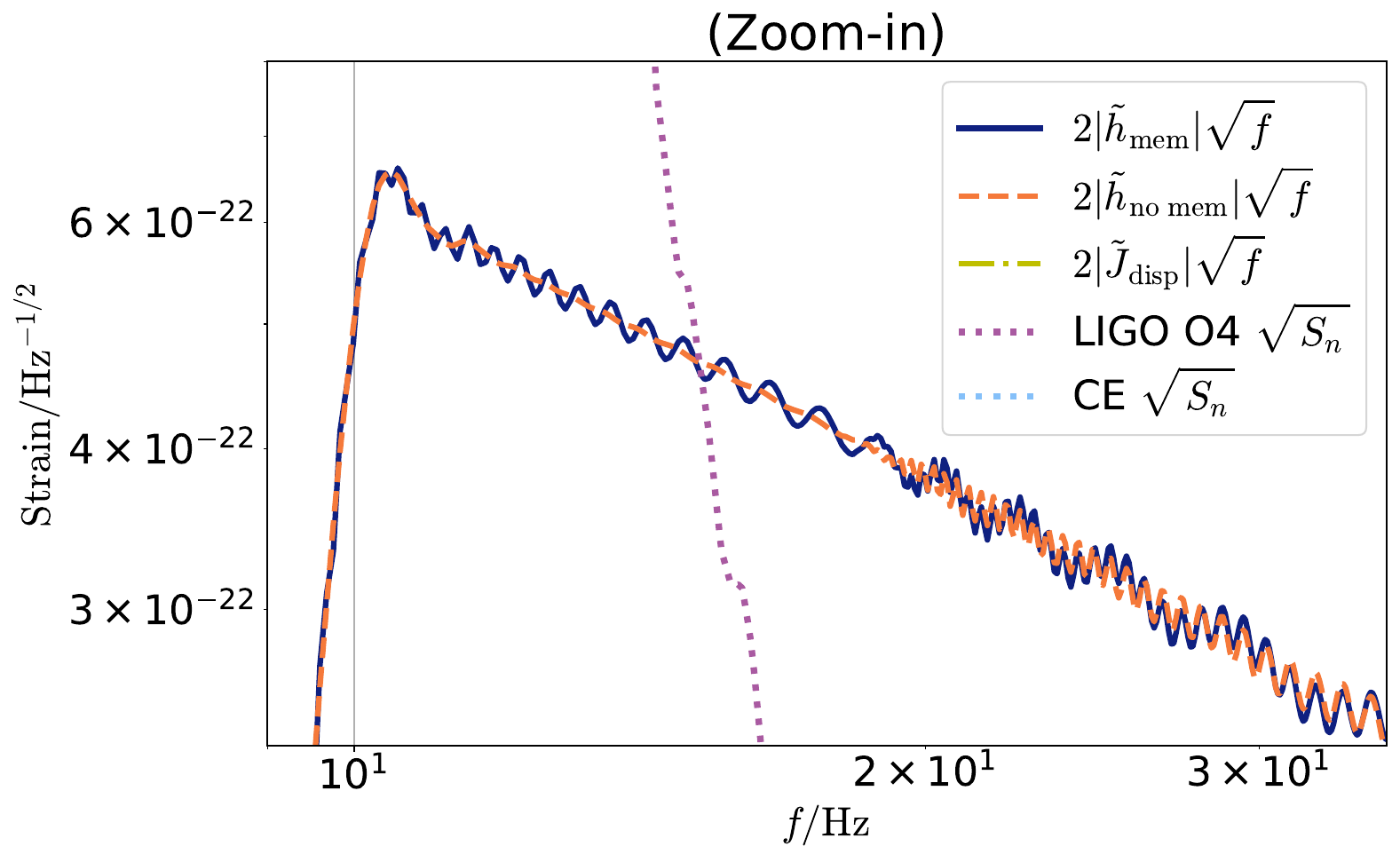}
	\caption{Spectra of simulated GW signals and noise curves of LIGO O4 and CE. The figure shows the frequency spectra of the memory signal $ h_{\text{mem}} $ (blue solid), the no-memory signal $ h_{\text{no mem}} $ (orange dashed), and the displacement memory $ J_{\text{disp}} $ (yellow dash-dotted), together with the detector noise PSD curves of the O4 LIGO (purple dotted) and CE (cyan dotted).  The signals are emitted from the BBH system simulated in Sec.~\ref{sec:bbh_simulation} with $ M_z=72M_\odot $ and $ d_{\mathrm{L}}=440 $~Mpc, viewed edge on from the Earth.}
	\label{fig:ligo_ce_asd}
\end{figure*}

In GW astronomy, the signal-to-noise ratio (SNR) is a common measure that quantifies the significance of a signal in a noisy detector.  This is essentially the square-root of the inner product from Eq.~\eqref{eq:inner_product} (with no time offset) of the signal with itself~\cite{Finn:1992wt}:
\begin{align}
	\rho = \sqrt{\langle s | s \rangle} = \left[4\int_{f_{\min}}^{f_{\max}} \frac{|\tilde{s}(f)|^2}{S_n(f)} df\right]^{1/2}. \label{eqn:snr}
\end{align}
The frequency-domain signal measured by a detector is related to the antenna pattern of the detector~\cite{Robson:2018ifk}:
\begin{align}
	\tilde{s}(f) &= F_+(f,\Theta, \Phi, \psi) \tilde{h}_+(f, \theta, \phi) \nonumber \\
	&\qquad + F_\times(f,\Theta, \Phi, \psi) \tilde{h}_\times(f, \theta, \phi). 
\end{align}
Here, $\tilde{h}_+$ and $\tilde{h}_\times$ are the Fourier transforms of $h_+$ and $h_\times$. The antenna-pattern response functions, $ F_+ $ and $ F_\times $, are generally complex valued and frequency dependent. They depend on the polarization angle $\psi$, a degree of freedom arising from rotating the detector about the line of sight. They also depend on the sky location $ (\Theta, \Phi) $, where $\Theta$ is the polar angle and $\Phi$ the azimuthal angle in the spherical coordinates of the detector.\footnote{A good diagram showing these angles can be found in \cite{Thorne1_skip} or \cite{Sathyaprakash:2009xs}. The angles $\Theta$ and $\Phi$ in this paper are denoted by $\theta$ and $\phi$ in these references.}

For ground-based detectors, we can assume the static limit (the long wavelength approximation) so that the response functions $ F_+ $ and $ F_\times $ are real and independent of frequency. The response functions for L-shaped interferometers, including LIGO and CE\footnote{While the static limit holds reasonably well for the current ground-based detectors (LIGO/Virgo/KAGRA), it starts to break down for the next-generation ground-based detectors. Keeping the assumption of the static limit may bring adverse impacts on GW analysis in the next-generation detectors (e.g., biased source localization of events detected by CE \cite{Essick:2017wyl}). For demonstration purposes, we still keep the static limit assumption for CE.}, can be found in \cite{Thorne1_skip, Sathyaprakash:2009xs}. In this section, we will compute two SNRs for each incoming GW in LIGO and CE. We calculate the first (second) SNR $ \rho_{\text{max}} $ ($ \rho_{\text{avg}} $) using the detector response [$ |\tilde{s}(f)|^2 $] that is maximized (averaged) over $ (\Theta, \Phi, \psi) $. For LIGO and CE, we have the relation $\rho_{\text{max}} = \sqrt{5} \rho_{\text{avg}}$, because $\langle F_+^2 \rangle = \langle F_\times^2 \rangle = 1/5$ and $\langle F_+F_\times \rangle = 0 $ \cite{Thorne1_skip}, where $\langle \cdot \rangle$ denotes the average over $ (\Theta, \Phi, \psi) $.

The signals studied in this section are generated from the same BBH source described in Sec.~\ref{sec:bbh_simulation}. We only consider the edge-on view of the BBH, $ (\theta=\pi/2, \phi=0) $, so the strain is $+$ polarized. We fix the (redshifted) total mass at $ M_z=72M_\odot $ and the distance at $ d_{\mathrm{L}}=440 $~Mpc. These values correspond to the median results from a reanalysis \cite{LIGOScientific:2018mvr} of the first detection event, GW150914, which differ slightly from the original analysis \cite{LIGOScientific:2016aoc}. The two black holes producing GW150914 had different masses and nonzero spins, but we believe our simulated BBH system (equal mass, no spin) with the total mass and distance of GW150914 is a also good representative of currently detectable BBH sources.

We consider GW signals both with and without memory. The memory signal, denoted by $ h_{\text{mem}} $, is merely $ h_+(t, \theta=\pi/2, \phi=0) $ with $ M_z=72M_\odot $ and $ d_{\mathrm{L}}=440 $~Mpc. With this particular total mass $M_z$, the signal $ h_{\text{mem}}$ (and the quantities $ h_{\text{no mem}} $ and $J_{\text{disp}}$ later introduced) has a duration of about 11.4~sec. We define the no-memory waveform as the ordinary part of the strain, $ J_m + J_{\widehat{N}} $. The no-memory signal, $ h_{\text{no mem}} $, is then $ \Re (J_m + J_{\widehat{N}}) $ evaluated at $\theta=\pi/2$, $\phi=0$, $ M_z=72M_\odot $, and $ d_{\mathrm{L}}=440 $~Mpc.

\begin{table}[t]
	\caption{References (second column) and sensitivity bands (third column) of the noise curves used in this paper.}
	\label{tbl:noise_curves}
	\begin{ruledtabular}
		\begin{tabular}{ccc}
			Detector & $S_n(f)$ & $ [f_{\min}, f_{\max}] $ (in Hz) \\
			\hline 
			LIGO O4 & \cite{LIGO_O4_noise_skip}\footnote{Used for O4 simulations in \cite{KAGRA:2013rdx}.} & $ [10, 4096] $ \\
			CE & \cite{CE_noise_skip}\footnote{The baseline configuration of a single 40~km CE interferometer. Used in \cite{Srivastava:2022slt}.} & $ [5, 4096] $ \\
			LISA & Eq.~(1) of \cite{Robson:2018ifk}\footnote{Using parameters for a 6-month mission. This noise curve incorporates the average antenna-pattern response functions, unlike LIGO's and CE's. Meanwhile, this curve includes the effects of LISA's three component interferometers that share arms mutually, while LIGO's and CE's curves assume the use of a single interferometer.} & [$10^{-5}$, 1]
		\end{tabular}
	\end{ruledtabular}
\end{table}

To calculate the SNR, Eq.~\eqref{eqn:snr}, we also need to specify the noise PSD curve. We consider two detectors, O4's LIGO and CE, whose noise PSD curves and sensitivity bands are listed in Table~\ref{tbl:noise_curves}. Figure~\ref{fig:ligo_ce_asd} shows the noise amplitude spectral density curve $ \sqrt{S_n(f)} $ of O4's LIGO (purple dotted) and CE (cyan dotted), together with the memory signal $ h_{\text{mem}} $ (blue solid) and the no-memory signal $ h_{\text{no mem}} $ (orange dashed).

We follow the plotting convention in \cite{Boyle:2009dg} and multiply the signal spectra by a factor of $ 2\sqrt{f} $, so that both the spectra and the noise curves share the same unit.  This normalization also lets us visually approximate the SNR based on the area between the signal and noise curves.  In the left panel of this figure, we see that both the memory and no-memory signals are well above the noise curves, which simply indicates the detectability of both signals. Visually, we expect that the memory effect will only be more significant than noise if the SNR of the total signal is greater than roughly 100, with significant low-frequency contributions. We calculate the SNRs of both signals in LIGO O4, CE, and LISA (more about LISA in Sec.~\ref{sec:snr_lisa_results}) and list them in Table~\ref{tbl:snr_numbers}. Note that for LIGO O4 and CE, we compute each SNR in a single interferometer, instead of a network of multiple interferometers.

\begin{table}[t]
	\caption{SNRs of the memory signal $ h_{\text{mem}} $, the no-memory signal $ h_{\text{no mem}} $, and the memory itself $J_{\text{disp}}$ detected by different interferometers. The interferometer is LIGO O4, CE, or LISA. Two types of SNRs, maximum SNR $ \rho_{\text{max}} $ and average SNR $ \rho_{\text{avg}} $, are computed. The signals are generated from the simulated BBH source at a distance of $d_{\mathrm{L}}=440$~Mpc. We use different initial total masses (second column) for different detectors when calculating SNRs.}
	\label{tbl:snr_numbers}
	\begin{ruledtabular}
		\begin{tabular}{cccccc}
			Detector & $M_z/M_\odot$ & SNR Type & $ h_{\text{mem}} $ & $ h_{\text{no mem}} $ & $J_{\text{disp}}$  \\
			\hline 
			LIGO O4 & 72 & maximum & 44.742 & 44.727 & 0.680\\
			& & average & 20.009 & 20.002 & 0.304\\
			CE & 72 & maximum & 1628.5 & 1627.9 & 40.3\\
			& & average & 728.29 & 728.01 & 18.04\\
			LISA & 4000 & average & 4.945 & 1.810 & 4.600
		\end{tabular}
	\end{ruledtabular}
\end{table}

In Fig.~\ref{fig:ligo_ce_asd}, we observe two major differences between $ \tilde{h}_{\text{mem}} $ and $ \tilde{h}_{\text{no mem}} $. Below the turn-on frequency at $ \sim 10 $~Hz, $ \tilde{h}_{\text{mem}} $ is much more powerful than $ \tilde{h}_{\text{no mem}} $. Since all oscillations are shut off in this range, the spectra must correspond to any low-frequency contents in the signals. For $ \tilde{h}_{\text{mem}} $, this content is the displacement memory, and we have already seen its low-frequency spectrum many times in Sec.~\ref{sec:preprocessing}. Another difference between the two signals is in the inspiral spectrum but hardly visible in the left panel, so we show the enlarged view in the right panel. The spectrum of $ h_{\text{mem}} $ (blue solid) contains additional oscillations compared to the spectrum of $ h_{\text{no mem}} $ (orange dashed). These oscillations have already been described in Sec.~\ref{sec:beats}, and they originate from the beating of the $ (2,\pm2) $ modes and the displacement memory in the $ (2,0) $ mode. The beating is absent in the $ h_{\text{no mem}} $ spectrum as it contains no displacement memory. Note that both the $ h_{\text{mem}} $ and $ h_{\text{no mem}} $ spectra contain small oscillations above $ \sim 20 $~Hz, which comes from the beating of the $ (2,\pm2) $ and $ (4,\pm4) $ modes (also see Sec.~\ref{sec:beats}). 

In the left panel of Fig.~\ref{fig:ligo_ce_asd}, we also include the frequency spectrum of the displacement memory (yellow dash-dotted), 
\begin{align}
	J_{\text{disp}}(t) = \Re J_{\mathcal{E}} (t, \theta=\pi/2, \phi=0),
\end{align}
for reference. Based on its relative position to the noise curves, we can qualitatively deduce that for a single LIGO-O4 interferometer, the displacement memory studied in this section is not detectable in a single event. Nevertheless, this memory is detectable for a single CE interferometer.  We will analyze the detectability of this memory quantitatively in Sec.~\ref{sec:result_param_infer}, using Bayesian inference. Before that, let us briefly examine a scenario where the memory component itself is more detectable than the oscillatory part in LISA.

\subsection{Detecting memory in LISA} \label{sec:snr_lisa_results}

LISA aims to detect GW signals with frequencies much lower than those of ground-based detectors---ranging from below $10^{-4}$~Hz to just above 0.1~Hz~\cite{LISA:2017pwj}.  This not only leads to observing very different signal characteristics for the various classes of compact binaries that may be detected, but also opens up more of the spectrum to accentuate the low-frequency memory.

Stellar-mass binaries evolve very slowly at these frequencies, spending years creeping through a very small portion of LISA's sensitive band before merging.  In this stage, because the low-frequency $(\ell,m)=(2,0)$ component will not have reached the LISA band, the beating described in Sec.~\ref{sec:beats} will not be present, and the memory component will be undetectable.  For a stellar-mass binary that will merge during LISA's lifetime, the oscillatory part of the signal would only be in the sensitive band of ground-based detectors, well beyond LISA's band.  But roughly simultaneously with the merger, the non-oscillatory memory would appear quite suddenly in the LISA band.  Unfortunately, because of the small amplitude of the memory given such low masses, it would be difficult to detect in LISA unless a merger happened to occur unusually close to Earth~\cite{Gerosa:2019dbe}.

\begin{figure}[t]
	\centering
	\includegraphics[width=\linewidth]{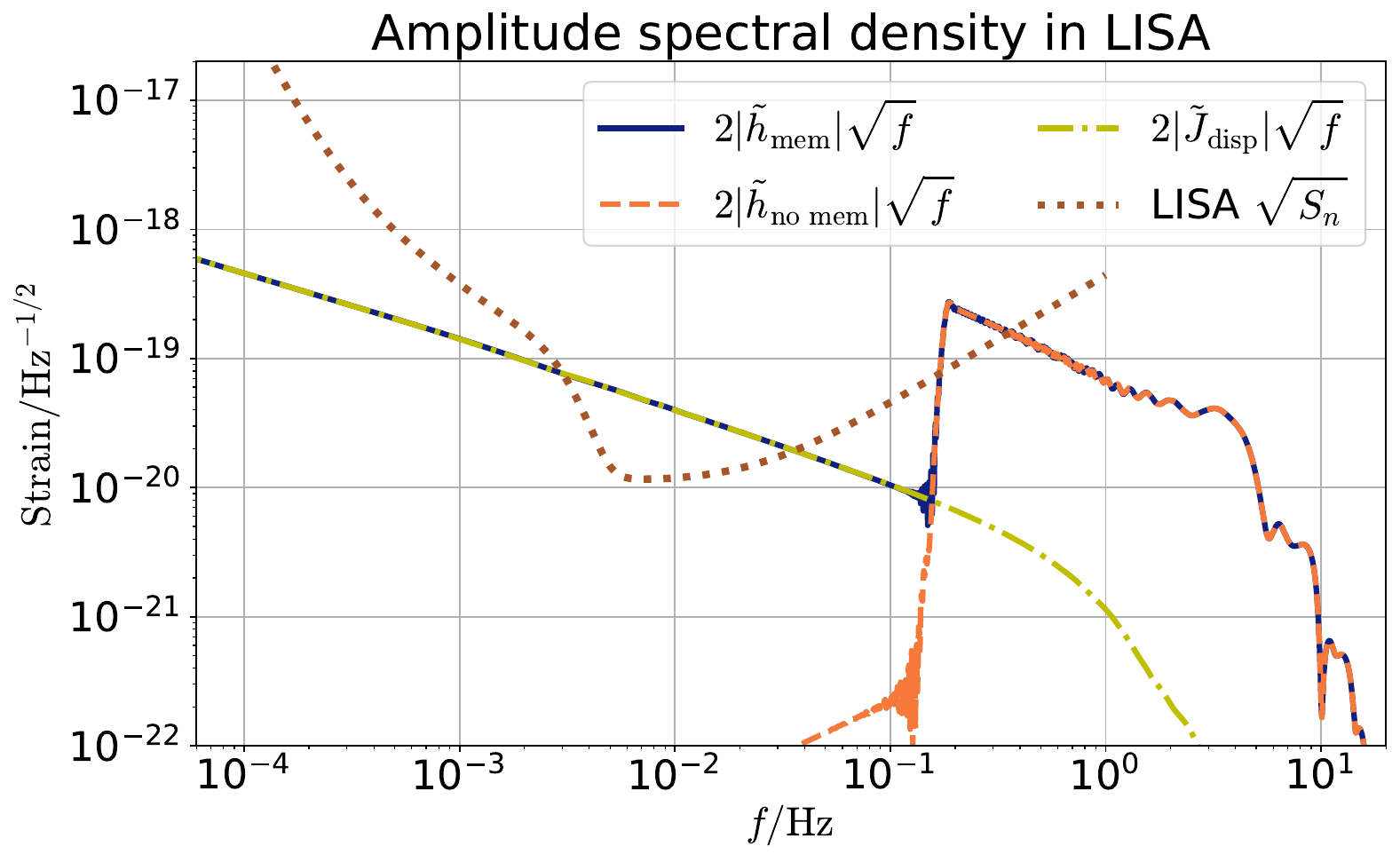}
	\caption{Spectra of simulated GW signals and LISA's noise curve. The figure shows the frequency spectra of the memory signal $ h_{\text{mem}} $ (blue solid), the no-memory signal $ h_{\text{no mem}} $ (orange dashed), and the displacement memory $ J_{\text{disp}} $ (yellow dash-dotted), together with the detector noise PSD curve of LISA (brown dotted; see Table~\ref{tbl:noise_curves} for its references). The signals originate from the edge-on view of the binary simulated in Sec.~\ref{sec:bbh_simulation} with $ M_z=4000M_\odot $ and $ d_{\mathrm{L}}=440 $~Mpc.}
	\label{fig:lisa_asd}
\end{figure}

Further up the mass scale, binaries will typically enter the LISA band closer to merger.  Supermassive binaries ($10^5M_\odot$ and higher) may merge within months of entering the sensitive band---or even hours for the most massive binaries.  For these, the basic detection picture is quite similar to that of the ground-based detectors.  Some intermediate-mass binaries (roughly $10^2$--$10^5\,M_\odot$) are likely to already be emitting within the LISA band when the mission first begins taking data.  In extreme cases, another possibility arises: the memory component of the signal may be detectable, while the oscillatory part may not be.  Figure~\ref{fig:lisa_asd} shows such an example, with plotting styles the same as in Fig.~\ref{fig:ligo_ce_asd}.  The total (redshifted) mass is $M_z = 4000\,M_\odot$.  Note, however, that we still choose the distance to be $440\,\mathrm{Mpc}$.  Though the uncertainties are still large, this is likely to be unusually close for such a massive binary~\cite{PortegiesZwart:2002iks, Berti:2006ew, Erickcek:2006xc, Barack:2018yly, Gerosa:2019dbe, Barausse:2020mdt}.  The binary is viewed edge on from Earth.  The distance and orientation are chosen to make the memory signal more visible.

Moreover, due to limitations of the numerical simulation, our example signal is significantly shorter than we would normally expect to find in LISA data.  At this mass, the simulated inspiral covers just $306.8\,\mathrm{sec}$.  At this distance, the SNR (averaged over sky location and polarization angle; see Table~\ref{tbl:snr_numbers}) of the memory component ($J_\text{disp}$) of the signal is $2.5$ times larger than the SNR of the oscillatory component ($ h_\text{no mem} $).  This circumstance suggests a method for detecting memory more directly.

To measure memory \emph{accurately}, we need to be confident that our model for the non-memory component of the waveform is accurate to a significantly greater degree than the memory effect itself.  There may be unmodeled effects such as the influence of matter in and around the binary, or degeneracies that make it impossible to precisely determine the correct oscillatory waveform.  More simply it may just not be practical to generate long template waveforms, as numerical simulations are too expensive and post-Newtonian approximations are not sufficiently accurate.  We can avoid these complications by taking only a relatively short piece of the data.  The oscillatory part of the signal can be calculated as usual, but is not present across most of the sensitive band, at the same time as the low-frequency memory component appears across the entire sensitive band.  In fact, with the post-processing procedure described in Sec.~\ref{sec:preprocessing}, the oscillatory component can easily be filtered out in the frequency domain, leaving only a distinct low-frequency memory component to be searched for in the data.

Taken to the extreme, this would leave us with limited data to exhibit any difference between the pre- and post-merger strain, and thus reduce the statistical power.  However, the inspiral portion of a signal measured for some time span $T$ at leading order~\cite{Damour:2000zb} will ``turn on'' at frequency $f_{2,2} \approx (5 M / 256 \nu T)^{3/8} / \pi M$.  Because of the $3/8$ exponent, measuring for longer increases the amount of non-memory content slowly. As long as $T$ is not \emph{too} large, this means that there will still be frequency separation between the two so that the higher-frequency component can still be filtered out effectively, while leaving enough power in the low-frequency component. The point is that there is a balance to be struck, and memory is one phenomenon that may actually push us to measure a \emph{shorter} signal.

\subsection{Detectability of memory in LIGO and CE} \label{sec:result_param_infer}

The powerful tool of Bayesian inference has been used extensively in all three observing runs \cite{LIGOScientific:2018mvr, LIGOScientific:2020ibl, LIGOScientific:2021usb, KAGRA:2021vkt} of the LIGO-Virgo-KAGRA collaboration. It serves multiple tasks in GW astronomy. One major task is parameter estimation, which is to construct the probability distributions of sky locations and source configurations. Another one is model comparison, which is to select a model against other models based on probabilistic evidence. Model comparison is our purpose of using Bayesian inference in this section.

The foundation of Bayesian inference is Bayes' theorem:
\begin{align}
	p(\theta|d, A) = \frac{\mathcal{L}(d|\theta, A) \pi(\theta|A)}{\mathcal{Z}(d|A)}. \label{eqn:Bayes}
\end{align}
Here, the prior distribution $ \pi(\theta|A) $ is the probability density of the parameter set $\theta$ assuming the model $A$. The parameter set is always associated with the model: a different model is in general specified by a different set of parameters. The posterior distribution $ p(\theta|d, A) $ is the probability density of $\theta$ given measured data $d$ and the model $A$. The likelihood function $ \mathcal{L}(d|\theta, A) $ is the probability density of $d$ given $A$ and $\theta$. The Bayes evidence 
\begin{align}
	\mathcal{Z}(d|A) = \int d\theta \mathcal{L}(d|\theta, A) \pi(\theta|A)
\end{align}
is the factor required to satisfy the normalization condition
\begin{align}
	\int d\theta p(\theta|d, A) = 1, 
\end{align}
and can be regarded as the probability of $d$ given $A$. 

Let $B$ be another model and its parameter set be $\nu$. The discussion in the previous paragraph still applies, except that we replace the symbols $A$ and $\theta$ by $B$ and $\nu$. Specifically, similar to the Bayes evidence $\mathcal{Z}(d|A)$ for the model $A$, we now have another Bayes evidence $\mathcal{Z}(d|B)$, for the model $B$. With these two evidences, we can construct the key metric in this section, the $A/B$ Bayes factor,
\begin{align}
	\text{BF}^A_B = \frac{\mathcal{Z}(d|A)}{\mathcal{Z}(d|B)},
\end{align}
which encodes the level of evidence in favor of one model over another. Rather than the Bayes factor itself, it is usually more convenient to work with its natural logarithm $ \log_e \text{BF}^A_B $. We use the convention in \cite{Lasky:2016knh, Thrane:2018qnx, Hubner:2019sly} that a threshold of $ |\log_e \text{BF}^A_B| = 8 $ marks the significance of this evidence level.

We simulate GW detection events and perform Bayesian inference using \texttt{Bilby} \cite{Ashton:2018jfp}.  We consider a LIGO-O4 or CE interferometer.  The noise curves and the sensitivity bands are listed in Table~\ref{tbl:noise_curves}. We choose a 4-second detection window, and truncate away any portion of the time-domain signal lying outside this window. We use two waveform templates in this study. One template contains memory and the other does not. The memory template uses both polarization components of the CCE waveform $ h $, while the no-memory template uses those of the ordinary part $ J_m + J_{\widehat{N}} $. Each template takes injection parameters as inputs and then outputs the frequency signals. The template has the following workflow. It intrinsically holds a time-domain numerical waveform whose unit of time is $M_z$. Based on the input total mass, we convert the waveform into a time-domain signal whose unit of time is second. Since we use a 4-second detection window, we keep the signal from $t=-2$~sec to $t=2$~sec and drop the portion outside this interval. As a reminder, $t=0$ is the merger time (see Sec.~\ref{sec:bbh_simulation}). Because the original numerical waveform has a finite length, we always choose an input total mass $M_z$ such that the inspiral signal is longer than 2 seconds. We also pad the ringdown signal to $t=2$~sec if necessary. We follow the ideas in Secs.~\ref{sec:tapering} and \ref{sec:line_subtraction} and preprocess the 4-second signal by first tapering the left side with a roll-on of 1 second\footnote{Here, we do not use the standard preprocessing scheme introduced in Sec.~\ref{sec:standard_procedure}.} and then line-subtracting the tapered signal. Finally, we apply the Fourier transform on the preprocessed signal and send the resulting frequency-domain signal to the template's output. 

We inject the GW signal into the detector using the memory template evaluated at the following injection parameters. We fix the total mass at $ M_z=72M_\odot $ and choose the optimal line of sight for memory detection: edge-on view of the binary (i.e., $ \theta=\pi/2 $ and $ \phi=0 $) and GW propagation perpendicular to the detector plane (i.e., $ \Theta=0 $). We also set the relative polarization angle $\psi$ to 0. Later, we will also specify the values of the distance used for injection.

\begin{table}[t]
	\caption{Injection parameters used for \texttt{Bilby}'s inference. Priors of the five inferred parameters are also included.}
	\label{tbl:param_infer}
	\begin{ruledtabular}
		\begin{tabular}{cc}
			Parameter & Value \\
			\hline 
			$\theta$ & $\pi/2$ \\
			$\phi$ & 0 \\
			Sampling frequency & 8192 Hz \\
			Signal duration & 4 sec \\
			Declination injected & 0 \\
			Declination prior & sine-uniform\footnote{Declination is uniform in sine. This prior is the same as the ``cos'' prior in \cite{Ashton:2018jfp}.} $ [-\pi/2, \pi/2] $ \\
			Right ascension injected & $ \alpha_0 \approx 1.828 $\footnote{We choose the value of the right ascension such that the detector plane is perpendicular to the propagation of the injected GW. This value depends on the geocentric time and the detector location.} \\
			Right ascension prior & uniform $ [\alpha_0-\pi, \alpha_0+\pi] $ \\
			$\psi$ injected & 0 \\
			$\psi$ prior & uniform $ [-\pi/2, \pi/2] $ \\
			$M_z$ injected & $ 72M_\odot $ \\
			$M_z$ prior & uniform $ [30M_\odot, 150M_\odot] $\footnote{The lower bound of the mass prior is limited by the finite length of the numerical waveform. It cannot be smaller than $ \sim 23M_\odot $ in this study. If the total mass is smaller than this value, the (padded) numerical waveform cannot cover the detection window of 4~sec. Hybridization with post-Newtonian waveforms can remedy this issue, but we choose not to implement it. This is because the injected signal is bright, and thus the mass posterior, which concentrates near the injected value, is insensitive to a wider prior.} \\
			$d_{\mathrm{L}}$ injected & 50~Mpc or 440~Mpc \\
			$d_{\mathrm{L}}$ prior & uniform $ [1~\text{Mpc}, 1000~\text{Mpc}] $ \\
		\end{tabular}
	\end{ruledtabular}
\end{table}

In this inference study, there are five parameters to be inferred: total mass $M_z$, distance $d_{\mathrm{L}}$, declination, right ascension, and polarization angle $\psi$. The prior distributions of these five parameters are listed in Table~\ref{tbl:param_infer}, together with the values of other injection parameters. For each template, \texttt{Bilby} constructs the posterior distributions for all five inferred parameters, along with the signal/noise $\log_e$ Bayes factor. By taking the difference between two templates' $\log_e$ Bayes factor, we obtain the memory/no-memory $\log_e$ Bayes factor, which we simply refer to as the $\log_e$ Bayes factor in the rest of this section. This $\log_e$ Bayes factor can be regarded as a random variable, whose randomness mainly\footnote{Another source of randomness comes from drawing ``live points'' in the nested sampling algorithm. These live points are random samples drawn from the prior distribution, and they are updated iteratively towards greater likelihoods \cite{Higson1_skip}.  We choose 2000 live points to keep this random effect minimal. Reweighting \cite{Payne:2019wmy} is another technique to reduce the live-point random effect.} comes from the detector noise. The detector noise, a random variable itself, is realized differently\footnote{Assume no random seed is used.} every time we inject a signal. Because of this randomness, a single realization of the $\log_e$ Bayes factor is not sufficient for our goal to compare two templates. Instead, we repeat the injection-inference simulations 100 times, each time with a new noise realization, and record the $\log_e$ Bayes factors. A distribution of the $\log_e$ Bayes factors is more reliable than any single value. We compare the distribution with the conventional threshold of $ \pm8 $ to claim any detectability of memory \textit{from a single event in a single detector}.

\begin{figure}[t]
	\centering
	\includegraphics[width=\linewidth]{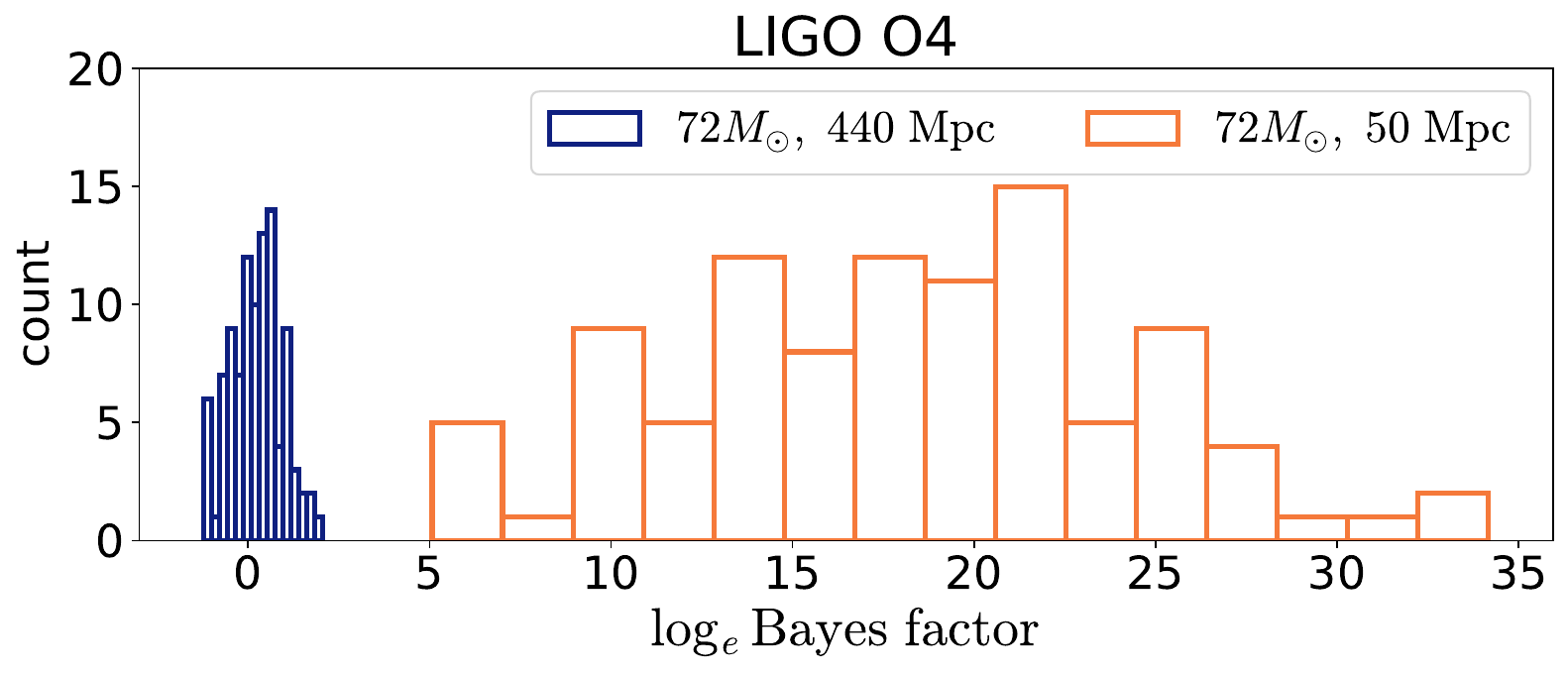}
	\includegraphics[width=\linewidth]{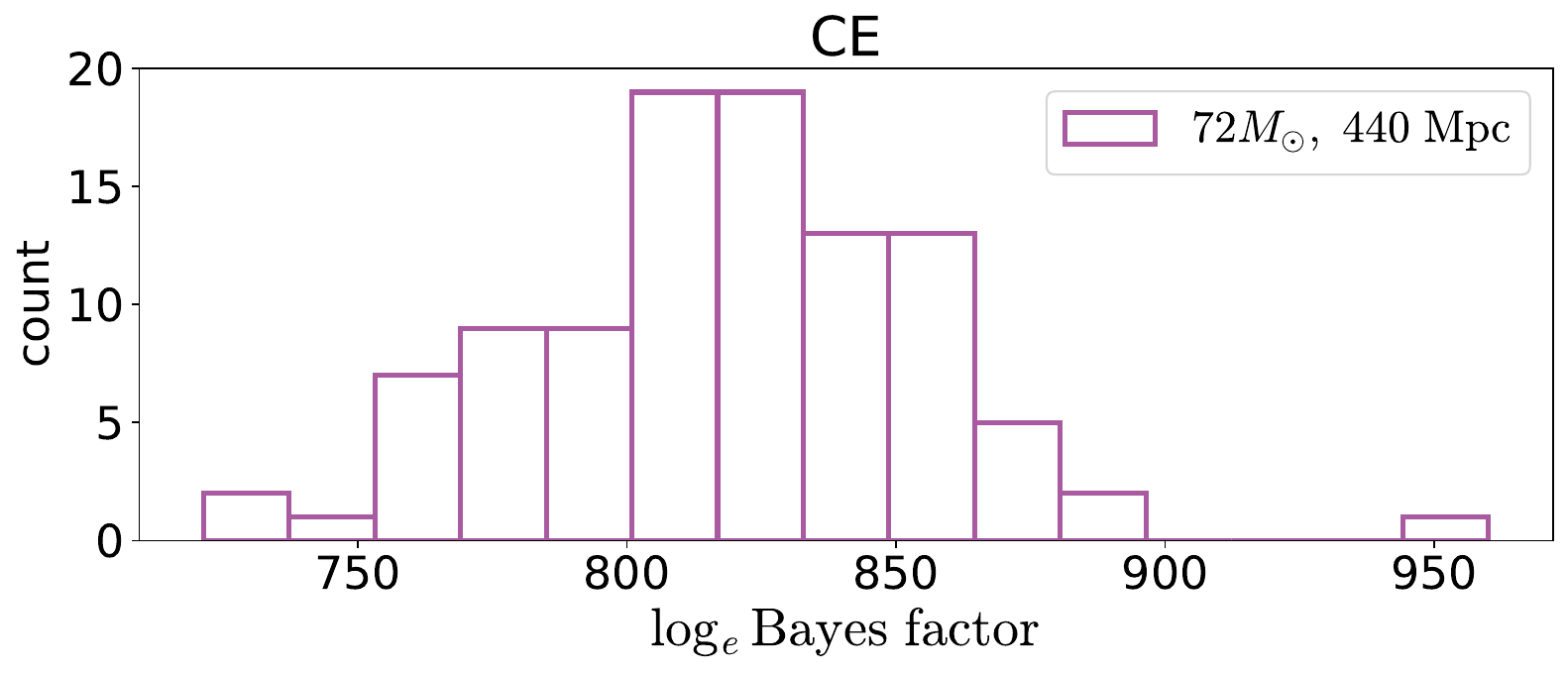}
	\caption{Histograms of the resulting $\log_e$ Bayes factors in this inference study. Three scenarios are considered: O4 LIGO detects a BBH system at a distance of 440~Mpc (blue) and another at 50~Mpc (orange), while CE detects a BBH at 440~Mpc (purple). All BBH sources have the same total mass of $ 72M_\odot $. Each scenario is simulated 100 times.}
	\label{fig:infer_bayes_factor_hist}
\end{figure}

We consider three detection scenarios. In the first scenario, we choose a LIGO-O4 detector and imagine a BBH source at a luminosity distance of 440~Mpc away from the Earth. The histogram of the resulting 100 $\log_e$ Bayes factors (because we simulate each scenario 100 times) is shown in blue in the top panel of Fig.~\ref{fig:infer_bayes_factor_hist}. The histogram is approximately normally distributed and is confined in a very narrow neighborhood of 0. Moreover, the whole histogram lies well within the conventional threshold of $ \pm8 $. This means that neither the memory template nor the no-memory template used in this study is more favorable than the other. In other words, the presence of the memory remains undetermined. Note that this scenario is very similar to the first GW event, GW150914. Thus, even with an edge-on BBH view (so that the memory is maximized), an upgraded LIGO detector, and the optimal sky location, we cannot detect the existence of memory in a GW150914-like signal.

In the next scenario, we keep using the same detector but bring the source closer from 440~Mpc to 50~Mpc. We plot the histogram of the $\log_e$ Bayes factors in orange in the top panel of Fig.~\ref{fig:infer_bayes_factor_hist}. We see that this histogram is symmetric but spread much more widely than the one in the first scenario. The majority of the distribution lies above the threshold of 8, which means that the memory template is favored over the no-memory template in most cases. Therefore, it is very likely that an O4 LIGO interferometer can detect the memory emitted by a BBH source at a luminosity distance of 50~Mpc away. However, 50~Mpc is a very short distance in GW astronomy. In fact, all BBH-associated detection events so far are located more than 100~Mpc away \cite{LIGOScientific:2018mvr, LIGOScientific:2020ibl, LIGOScientific:2021usb, KAGRA:2021vkt}. Only if we are very lucky can a BBH merger take place at such a short distance and lead to memory detection.

In the last scenario, we bring the source back to the distance at 440~Mpc but replace the LIGO detector by a CE detector. We show the $\log_e$-Bayes-factor histogram in purple in a separate panel (bottom) of Fig.~\ref{fig:infer_bayes_factor_hist}, because the $\log_e$ Bayes factors are distributed far away from the other two histograms in the previous scenarios. This is not surprising, because after all, CE is significantly more sensitive than LIGO, by several orders of magnitude (see the left panel of Fig.~\ref{fig:ligo_ce_asd}). Hence, detecting memory will be a simple task for the next-generation detectors.

\section{Conclusion} \label{sec:conclusion}

In this paper, we re-examine the preprocessing procedures that are applied to time-domain numerical waveforms before we Fourier-transform them to the frequency domain. The sample gravitational waveform studied throughout this paper is extracted from an equal-mass non-spinning BBH simulation. With the help of the CCE method, this waveform contains expected memory effect. We find that the common preprocessing scheme in literature, which consists of padding a signal by zeros and tapering both of its sides with a window function, does not always produce the correct waveform spectrum in the low-frequency range. In particular, the frequency spectrum of the displacement memory preprocessed using the common scheme is inconsistent with the zero-frequency limit of the memory. The spectral leakage induced by the step of windowing contaminates the low-frequency spectrum, which can potentially impair any memory-related researches, e.g., detecting memory in interferometers. To address this issue, we propose a new three-step preprocessing scheme: appending constant values to the time-domain signal, tapering the inspiral part of the signal, and subtracting a line connecting both ends from the padded tapered signal. We confirm that the frequency spectrum of the displacement memory preprocessed using this new scheme indeed approaches the zero-frequency limit. We also demonstrate the robustness of the new scheme by varying the parameters used in preprocessing, such as the padding and tapering lengths. 

As an application of this new preprocessing scheme, we investigate several characteristics in the frequency spectrum of a GW signal with memory. We simulate this signal by evaluating the numerical waveform along the edge-on direction of the BBH, so that the displacement memory effect is maximized. We find that the high-frequency portion of the spectrum exhibits a stepped decreasing pattern, which is made of not only the $ (\ell, m)=(2,2) $ mode but also higher even-numbered $ \ell =m $ modes, e.g., the $ (4,4) $, $ (6,6) $, and $ (8,8) $ modes. We also find two beating patterns that are visible in the full spectrum. One is formed by the dominant and subdominant oscillatory modes, while the other by the dominant oscillatory mode and the displacement memory. 

Because the memory leaves an imprint on the frequency spectrum, it is potentially observable. We explore the possibilities of detecting the memory in both current and future interferometers using the methods of Bayesian inference. We find that the memory remains undetectable in a single event for the current LIGO, unless the BBH source is incredibly close to the Earth. In contrast, the next-generation detector CE, which is considerably more sensitive than LIGO, has a much greater potential in detecting the memory in a single event.  We discuss a detection scenario in LISA where memory is more detectable than oscillations from an intermediate-mass BBH.  We also calculate the SNRs of sample GW signals with and without memory in LIGO, CE, and LISA. 

As future work, one can apply the new preprocessing scheme to waveforms of more generic BBH configurations, such as high-mass-ratio, precessing, or eccentric systems. A larger collection of waveform spectra from general BBH systems possibly offers us a better understanding of the low-frequency behaviors in the waveforms. Regarding the detectability of the memory, one can combine an ensemble of events, as in \cite{Lasky:2016knh, Hubner:2019sly, Boersma:2020gxx, Hubner:2021amk, Grant:2022bla}, to estimate the number of events (or years) required to confirm the existence of the displacement memory. With the new scheme, such a study can be performed more accurately.

\section*{Acknowledgments}

We thank Jooheon Yoo for useful discussions. Computations for this work were performed with the Wheeler cluster at Caltech, the Bridges-2 system at the Pittsburgh Supercomputing Center (PSC), and the Frontera system at the Texas Advanced Computing Center (TACC). This work was supported in part by the Sherman Fairchild Foundation and by NSF Grants PHY-2207342 and OAC-2209655 at Cornell.

\newpage
\bibliographystyle{apsrev4-2}
\bibliography{library,library2}
\end{document}